\documentclass[11pt]{article}
\usepackage{jheppub}

\makeatletter
\def\@fpheader{\relax}
\makeatother

\usepackage{amstext,amsmath,amssymb,amsfonts,bbm, amsthm}
\usepackage{graphicx}
\usepackage{hyperref}
\usepackage{cases}
\usepackage{tocvsec2}

\settocdepth{subsection}

\def\beq{\begin{equation}}
\def\be{\begin{equation}}
\def\ee{\end{equation}}
\def\bes{\begin{eqnarray}}
\def\ees{\end{eqnarray}}

\def\f{\frac}

\def\pp{\partial}

\DeclareMathOperator\arctanh{tanh^{-1}}
\DeclareMathOperator\arccoth{coth^{-1}}

\newcommand{\scri}{\mathcal{I}}

\begin{document}

\title{Entanglement entropy production in gravitational collapse: covariant regularization and solvable models}

\author[1]{Eugenio Bianchi,\note{ebianchi@gravity.psu.edu}}
\author[2]{Tommaso De Lorenzo\note{tommasodelorenzo@yahoo.it}}
\author[3]{and Matteo Smerlak\note{msmerlak@perimeterinstitute.ca}}

\affiliation[1]{Institute for Gravitation and the Cosmos \& Physics Department,\\ Penn State, University Park, PA 16802, USA}
\affiliation[2]{Universit\`a di Pisa, Dipartimento di Fisica ``Enrico Fermi'', \\Largo Bruno Pontecorvo 3, 56127 Pisa, Italy}
\affiliation[3]{Perimeter Institute for Theoretical Physics,\\ 31 Caroline St.~N., Waterloo ON N2L 2Y5, Canada}

\vspace{1em}

\date{\small\today}

\abstract{We study the dynamics of vacuum entanglement in the process of gravitational collapse and subsequent black hole evaporation. In the first part of the paper, we introduce a covariant regularization of entanglement entropy tailored to curved spacetimes; this regularization allows us to propose precise definitions for the concepts of black hole ``exterior entropy'' and ``radiation entropy.'' For a Vaidya model of collapse we find results consistent with the standard thermodynamic properties of Hawking radiation. In the second part of the paper, we compute the vacuum entanglement entropy of various spherically-symmetric spacetimes of interest, including the nonsingular black hole model of Bardeen, Hayward, Frolov and Rovelli-Vidotto and the ``black hole fireworks'' model of Haggard-Rovelli. We discuss specifically the role of event and trapping horizons in connection with the behavior of the radiation entropy at future null infinity. We observe in particular that $(i)$ in the presence of an event horizon the radiation entropy diverges at the end of the evaporation process, $(ii)$ in models of nonsingular evaporation (with a trapped region but no event horizon) the generalized second law holds only at early times and is violated in the ``purifying'' phase, $(iii)$ at late times the radiation entropy can become negative (i.e. the radiation can be less correlated than the vacuum) before going back to zero leading to an up-down-up behavior for the Page curve of a unitarily evaporating black hole.}

\maketitle
\flushbottom

\section{Introduction}
In quantum field theory the existence of correlations at spacelike separations indicates that the vacuum is a highly entangled state. Entanglement entropy provides a measure of these correlations \cite{Sorkin:2014kta,Bombelli:1986rw,Srednicki:1993im}. When the vacuum state is perturbed, for instance because of the coupling to an external background field, the amount of entanglement in the vacuum can change. In this paper we study the evolution of the entanglement entropy of the vacuum due to the coupling to an external gravitational field describing the collapse of classical matter.

The entanglement entropy is generally defined---at a given time---as a measure of the entanglement between modes of the field supported respectively in a spatial region and its complement \cite{Sorkin:2014kta,Bombelli:1986rw,Srednicki:1993im}. In a general-relativistic setting, it is desirable to have a covariant definition associated to spacetime regions (or causal domains) instead of regions of space at a given time \cite{Sorkin:2012sn,Czech:2012bh,Bianchi:2012ev}. This is in fact possible thanks to the causality and unitarity properties of relativistic quantum field theory \cite{haag1996local}. In the first part of this paper we adopt this spacetime perspective and introduce a covariant regularization of the entanglement entropy of causal domains, the \emph{causal-splitting regularization} (Sec.~\ref{sec:regularization}). The regularized entropy is defined in terms of the mutual information of causal domains separated by a splitting region \cite{Casini:2008wt,werner1987local,Yngvason:2014oia}. The covariant cut-off is given by the spacetime volume of the splitting region. This covariant definition of the entanglement entropy is tailored to curved spacetimes; it allows us to compare the entanglement entropy of different spacetime regions and define a cut-off independent notion of \emph{entanglement entropy production}.

In Sec.~\ref{sec: gravitational collapse} we consider the entanglement entropy production during gravitational collapse. A spherically symmetric distribution of classical matter collapses and forms a star or a black hole. At past null infinity $\scri^-$ the background geometry is asymptotically flat and a massless quantum field is prepared in the in-going vacuum state. The quantum field is treated as a test field, with no backreaction on the geometry. We restrict our attention to spherically symmetric modes of the field ($s$-wave) and neglect backscattering (geometric optics approximation). These assumptions reduce the analysis to a two-dimensional quantum field theory problem \cite{Birrell:1982ix,Fabbri:2005mw} and classical results in conformal field theory \cite{Holzhey:1994we} can be used to compute the entanglement entropy production.

We consider the time evolution of the entanglement entropy for three different regions of spacetime: $(i)$ a thick-shell region far from the collapsed object, $(ii)$ the exterior of the event horizon when a black hole forms, and $(iii)$ a portion of future null infinity. In all three cases the entanglement entropy production can be connected to the Hawking process and to its thermodynamics: the propagation of the quantum field in a gravitational-collapse spacetime in general results in the production of radiation, i.e. of an excited state of the field at future null infinity $\scri^+$. To illustrate this relation, in Sec.~\ref{sec: Vaidya} we study in detail the entanglement entropy production for a black hole formed by the collapse of a thin shell (Vaidya spacetime).


In Sec.~\ref{sec: radiation} we study the entanglement entropy of the radiation emitted at future null infinity $\mathcal{I}^+$ (the ``Page curve'') for four analytically solvable toy models of gravitational collapse: the formation of a compact star, the formation and evaporation of a black hole with event horizon, the formation and evaporation of a non-singular black hole with a closed trapped region but no event horizon, and the tunnelling of a black hole to a white hole. In particular we discuss some unexpected features of the Page curve relevant for the puzzle of information loss.

\section{Entanglement entropy of causal domains}
\label{sec:regularization}

In this section, we introduce a covariant regularization of vacuum entanglement entropy tailored to curved spacetimes. This regularization is based on the notion of mutual information of disconnected causal domains. For two-dimensional conformal fields, a formula of Holzhey, Larsen and Wilczek  \cite{Holzhey:1994we} permits explicit computations of the regularized entropy and allows us to define a cut-off independent notion of entanglement entropy production.\\

\subsection{General definitions}

Consider a $(d+1)$-dimensional (globally hyperbolic) spacetime with metric $g_{\alpha\beta}$, and let $S$ be a set of points. The causal complement $\overline{S}$ of $S$ is the set of all points which are space-like separated from all points of $S$. A causal domain $D$ is defined as a causally complete set, i.e. $D=\overline{\overline{D}}$. Given a Cauchy surface $\Sigma$ and a spatial region $R\subset \Sigma$, the Cauchy development of $R$ defines a causal domain $D=\mathcal{D}(R)\equiv \mathcal{D}^+(R)\cup \mathcal{D}^-(R)$ \cite{Wald:1984rg}. The causal complement of $D$ coincides with the Cauchy development of the complementary region in $\Sigma$, i.e. $\overline{D}=\mathcal{D}(\Sigma-R)$. We define the corner of the causal domain $D$ as $C_D\equiv \partial R$. Clearly every spatial region $R$ with the same boundary $C$ defines the same causal domain.\footnote{The causal domain with corner $C$ is defined by the intersection of the causal complements of each point in $C$, i.e. $D=\cap_{p\in C} \,\overline{p}$. No reference to the spatial region $R\subset\Sigma$ is needed.}

Given a pure global state $\rho$ in a quantum field theory, the entanglement entropy of the causal domain $D$ is usually defined by introducing a $UV$ cut-off $\epsilon$ and computing the von Neumann entropy of the reduced state $\rho_D$,\footnote{The reduced state $\rho_D$ is defined in terms of the global state $\rho$ and the local algebra of operators with support in $D$, see \cite{haag1996local}.}
\begin{equation}
S_\epsilon(D)=-\text{Tr} (\rho_D \log \rho_D)\,.
\label{eq:Seps}
\end{equation}
This quantity provides a measure of the correlations between the causally disconnected domains $D$ and $\overline{D}$. It diverges in the limit $\epsilon \to 0 $ due to the presence of $UV$ correlations in the state $\rho$.

\begin{figure}
\begin{center}
\includegraphics[width=0.6\textwidth]{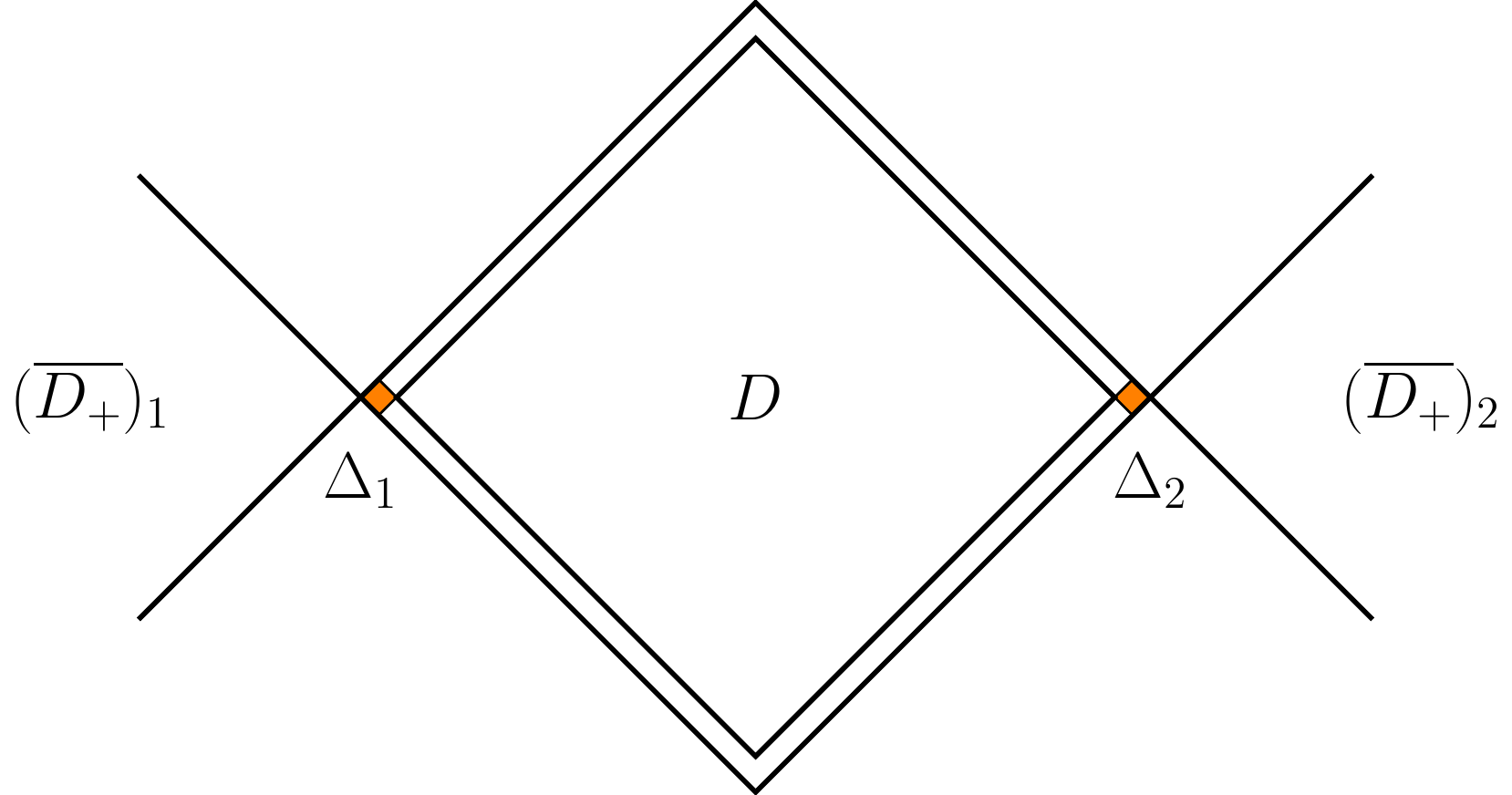}
\end{center}
\caption{Entanglement entropy in the \emph{causal splitting regularization} of a diamond $D$, defined as (one half) the mutual information between $D$ and $\overline{D_+}=(\overline{D_+})_{1}\cup(\overline{D_+})_{2}$. The covariant cutoff $\mu$ is the spacetime volume of the splitting region $\Delta=\Delta_{1}\cup\Delta_{2}$ (shaded).}
 \label{smearing}
        \end{figure}

\subsection{Causal splitting regularization}

Consider now two causal domains $D_1$ and $D_2$ that are disjoint, $D_1\cap D_2=\emptyset$, and causally disconnected, i.e. $D_1\subset \overline{D_2}$. The mutual information $I(D_1,D_2)$ of the two causal domains is defined as the relative entropy of the reduced state $\rho_{D_1\cup D_2}$ with respect to the tensor product of reduced states $\rho_{D_1}\otimes\rho_{D_2}$,
\begin{align}
I(D_1,D_2)\equiv&\;S(\rho_{D_1\cup D_2}|\rho_{D_1}\!\otimes\rho_{D_2})\, \label{eq:mutual-ax}\\[5pt]
=&\;\text{Tr}(\rho_{D_1\cup D_2} \log\rho_{D_1\cup D_2}\,-\rho_{D_1\cup D_2} \log \rho_{D_1}\!\otimes\rho_{D_2})\nonumber
\end{align}
The relative entropy is a well-defined quantity in quantum field theory, no $UV$ cut-off is required in its definition\footnote{The abstract definition of relative entropy in terms of von Neumann algebras can be found in \cite{araki1975relative}. See also ch. II of \cite{ohya2004quantum} for a pedagogical introduction.}. We now introduce a regularization $S_+(D)$ of the entanglement entropy that makes use of the notion of mutual information \cite{Casini:2008wt}.

Let $D_+$ be a causal domain that contains $D$. We call $D_+$ a smearing of $D$ and we are interested in the limit $D_+\to D$. The domains $D$ and $\overline{D_+}$ are causally disconnected. We define the splitting region as 
\begin{equation}
\Delta \equiv\overline{D\cup\overline{\!D_+\!}\;}\,.
\end{equation}
The splitting region $\Delta$ is causally complete and therefore is also a causal domain, see Fig.~\ref{smearing}.

At every point of the corner $C_D$ of the domain $D$ there are two null geodesics $\ell=\partial_v$ and $n=\partial_w$ that lie on the boundary of $\Delta$. In the limit $D_+\to D$ the spacetime volume $\mathcal{V}_{(d+1)}$ of $\Delta$ is given by the integral over $C_{\Delta}$ of the transversal spacetime area of $\Delta$, i.e.
\be
\mu\equiv g_{\alpha\beta}\ell^\alpha n^\beta\,\delta v\,\delta w. 
\ee
We require the smearing $D_+$ to be such that the transversal area $\mu$ is constant. As a result the splitting region has finite spacetime volume given by
\begin{equation}
\mathcal{V}_{(d+1)}(\Delta)\,=\;\mu\,\mathcal{A}_{(d-1)}(C_D)\,,
\label{eq:mu}
\end{equation}
where $\mathcal{A}_{(d-1)}$ is the area\footnote{For $d=1$ we define $\mathcal{A}_{(0)}=1$ and impose $\mathcal{V}_{(1+1)}(\Delta_i)=\mu$ for each connected component of $\Delta$.} of the $(d-1)$-dimensional corner $C_D$, and $\mu$ is a cut-off with dimensions of \emph{length}$\times$\emph{time}.

The entanglement entropy in the \emph{causal-splitting regularization} $S_+(D)$ is defined as half of the mutual information between the domain $D$ and the complement of its smearing $D_+$,
\begin{equation}
S_+(D)\equiv\frac{1}{2}\,I(D,\;\overline{\!D_+\!\!}\;)\,.
\end{equation}
For a finite cut-off $\mu$ this quantity is finite. In the limit $\mu\to 0$ the causal domains $D$ and its smearing $D_+$ coincide and the mutual information diverges. The point of view adopted in this paper is that $\mu$ is a physical cut-off, fixed for instance at the Planck scale
\begin{equation}
\mu=\frac{G\hbar}{c^4}\,,
\end{equation}
or at the scale below the point where the effective field theory considered breaks down. The cut-off is defined in a covariant way by the curved spacetime volume of the splitting region $\Delta$, formula (\ref{eq:mu}).

Now we connect the expression of the entanglement entropy $S_+(D)$ defined by the causal-splitting regularization to the standard formula (\ref{eq:Seps}).  Introducing the $UV$ cut-off $\epsilon$ and using the formula $S(\rho|\sigma)=\text{Tr}(\rho \log\rho\,-\rho \log \sigma)$ for the relative entropy, the mutual information (\ref{eq:mutual-ax}) can be written as \cite{vedral2002role}
\begin{equation}
I(D_1,D_2)=S_\epsilon(D_1)+S_\epsilon(D_2)-S_\epsilon(D_1\cup D_2)\,,
\end{equation} 
where a limit $\epsilon\to 0$ is understood on the right-hand side of the equation. Clearly, the mutual information remains finite in this limit. Using the fact that for a pure global state $S_\epsilon(D)=S_\epsilon(\overline{D})$, we find a simple expression for the entanglement entropy defined by the causal splitting:
\begin{equation}
S_+(D)=\frac{1}{2}\Big(S_\epsilon(D)+S_\epsilon(D_+)-S_\epsilon(\Delta)\Big)\,.
\label{eq:split}
\end{equation}
This expression contains two cut-offs, $\mu$ and $\epsilon$. The entropy $S_+(D)$ is defined by $\epsilon\to 0$ with $\mu$ finite. In the opposite limit, $\mu\to 0$ and $\epsilon$ finite, we have $S_\epsilon(\Delta)\to 0$, and the right-hand-side of (\ref{eq:split}) reduces to the ordinary entropy $S_\epsilon(D)$.

Defining the entanglement entropy in terms of mutual information of complementary regions has various advantages, especially in the presence of gauge fields \cite{Casini:2013rba}. In this paper we are mostly interested in its use in a curved background spacetime where the causal-splitting regularization allows us to compare the entanglement entropy of different regions of spacetime while keeping the same physical cut-off $\mu$ constant.


\subsection{Entanglement entropy in two-dimensional spacetimes}


In $(1+1)$-dimensional Minkowski space, a causal domain (or diamond) is determined by two spacelike separated points, the corners of the diamond: $p_1=(v_1,w_1)$ and $p_2=(v_2,w_2)$ with $v_2<v_1$ and $w_1<w_2$. Here $v$ and $w$ are inertial null coordinates, the metric is $ds^2=-dv\,dw$, and the causal domain is the set $D=[v_2,v_1]\times[w_1,w_2]$. The standard expression of the  entanglement entropy of a massless scalar field in the Minkowski vacuum is \cite{Callan:1994py,Holzhey:1994we,Calabrese:2004eu}\begin{equation}
S_\epsilon(D)=\frac{1}{6}\log \frac{\Delta v\,\Delta w}{\epsilon^2}\,,
\label{eq:Holzhey}
\end{equation}
where $\Delta v\equiv v_1-v_2$, $\Delta w\equiv w_2-w_1$, and $\epsilon$ is an ultraviolet cut-off.\footnote{This formula is most easily derived using Euclidean path integral methods, with the cut-off $\epsilon$  corresponding to a smearing of the conical defects \cite{Holzhey:1994we}. The formula can also be derived using real time methods by imposing a cut-off on field modes as a Dirichlet condition at distance $\epsilon$ from the boundary \cite{Callan:1994py}, or by introducing a lattice regularization with lattice spacing $\epsilon$ \cite{Casini:2009sr}. It generalizes to any two-dimensional conformal field theory \cite{Holzhey:1994we,calabrese2009entanglement}.} We now introduce a smearing in the size of the diamond $D$. In particular we consider a larger diamond $D_+= [v_2-\delta v_2,v_1+\delta v_1]\times [w_1-\delta w_1,w_2+\delta w_2]$, with $\delta v_1, \delta v_2, \delta w_1, \delta w_2$  all positive. The causal complement of $D\cup\overline{\!D}_+$ is a domain $\Delta$ consisting of two (small) disconnected diamonds,
\begin{equation}
\Delta=\Delta_1\cup \Delta_2\,,
\end{equation}
with $\Delta_1=[v_1,v_1+\delta v_1]\times [w_1-\delta w_1, w_1]$ and $\Delta_2=[v_2-\delta v_2,v_2]\times [w_2,w_2+\delta w_2]$, see Fig.~\ref{smearing}. 

The entanglement entropy defined via a causal splitting can be computed using formula (\ref{eq:split}), the expression (\ref{eq:Seps}), and the fact that the entanglement entropy of the union of two diamonds is additive in the limit of diamonds that are small compared to their separation, $S(\Delta_1\cup \Delta_2)\to S(\Delta_1)+S(\Delta_2)$ \cite{Casini:2008wt,Calabrese:2009ez}. This results in the expression
\begin{equation}
S_+(D)=\frac{1}{12}\log\frac{(\Delta v)^2 (\Delta w)^2}{\delta v_1\, \delta w_1	\, \delta v_2 \, \delta w_2}\,.
\label{eq:smear2d}
\end{equation}
To conclude the derivation of the entanglement entropy in the causal-splitting regularization, the spacetime volume of the splitting regions $\Delta_1$ and $\Delta_2$ must now be expressed in terms of the physical cut-off $\mu$, namely $\delta v_1 \delta w_1=\delta v_2 \delta w_2=\mu$. As a result, the spacetime volume cut-off $\mu$ takes the place of the $UV$ cut-off $\epsilon^2$ in (\ref{eq:Holzhey}). Notice that the cut-off $\mu$ is Lorentz invariant: under a boost the shape of the splitting region changes, $\delta v_1\to \lambda\, \delta v_1$ and $\delta w_1\to \lambda^{-1} \delta w_1$, but its spacetime volume $\delta v_1 \delta w_1=\mu$ remains invariant.


\subsubsection{Two-dimensional curved spacetimes}

The relation between the causal-splitting regularization and the standard regularization of the entanglement entropy becomes non-trivial in a curved spacetime. Consider a spacetime with metric
\begin{equation}
ds^2=- C^2(v,w)\,dv\,dw
\label{eq:Omegauv}
\end{equation}
and the same past asymptotic structure as Minkowski space. A minimally coupled massless scalar field on this curved background satisfies the wave equation $0=\Box \,\varphi\,=\, C^{-2}\,\partial_v\partial_w \varphi$. Therefore in terms of the coordinates $v$ and $w$ the solutions of the wave equation are the same as in Minkowski space and the global state $\rho$ defined by the Minkowski vacuum at past null infinity is also a global state of the quantum theory on the curved spacetime (\ref{eq:Omegauv}).\footnote{This argument generalizes to conformal vacua in any conformal field theory.} As a result, its entanglement entropy has the same expression (\ref{eq:smear2d}) as in Minkowski space. What changes now is the metric relation between the points $(v_1,w_1)$ and $(v_2,w_2)$, and most importantly the relation between the splittings $\delta v_1, \delta v_2, \delta w_1, \delta w_2$, and the covariant cut-off $\mu$ given by the volume of the splitting region
\begin{equation}
- C^2(v_1,w_1)\,\delta v_1 \delta w_1\,=\,- C^2(v_2,w_2)\,\delta v_2 \delta w_2\,=\,\mu\,.
\end{equation}
Thus, in the causal-splitting regularization, the expression of the entanglement entropy of a causal domain $D$ with corners $p_1=(v_1,w_1)$ and $p_2=(v_2,w_2)$ in a curved spacetime is given by
\begin{equation}
S_+(D)=\frac{1}{12}\log\!\frac{(\Delta v)^2 (\Delta w)^2\, C^2(v_1,\!w_1)\, C^2(v_2,\!w_2)}{\mu^2}\,.
\label{eq:Smu}
\end{equation}


\subsubsection{Entanglement entropy production}

These preliminaries allow us to address the main objective of this paper, namely computing the entanglement entropy \emph{production} in the Hawking process. A massless scalar field prepared in the Minkowski vacuum at $\mathcal{I}^-$ evolves in the time-dependent background describing a gravitational collapse, and at $\mathcal{I}^+$ is found in an excited state. The dynamics of the background results in particle production \cite{Hawking:1974sw}. We wish to probe the evolution of the state of the field and the emission of Hawking radiation by studying the evolution of the entanglement entropy of the field.

For this purpose, let us consider a one-parameter family of diamonds $D_\lambda$ labeled by the trajectory of the two space-like separated corners $p_1(\lambda)=\big(v_1(\lambda),w_1(\lambda)\big)$ and $p_2(\lambda)=\big(v_2(\lambda),w_2(\lambda)\big)$. Given a reference `time' $\lambda_{0}$, we define the \emph{entanglement entropy production}, or \emph{excess entanglement entropy}, in $D_{\lambda}$ as 
\be
\Delta S(\lambda)\equiv\,S_{+}(D_{\lambda})-S_{+}(D_{\lambda_{0}})\,.
\ee
Recalling that $\mu$ is a physical cut-off that is kept \emph{fixed} in the evolution, we can compute the entanglement entropy production using formula \eqref{eq:Smu} and find the $\mu$-independent result
\begin{equation}
\Delta S(\lambda)=\frac{1}{12}\log\frac{(\Delta v)^2(\Delta w)^2\, C_1^2\;  C_2^2\big\vert_{\lambda}}{(\Delta v)^2 (\Delta w)^2\,\, C_1^2\;  C_2^2\big\vert_{\lambda_{0}}}\,.
\label{mainformula}
\end{equation}
where $\Delta v\vert_{\lambda}\equiv v_1(\lambda)-v_2(\lambda)$, $\,\Delta w\vert_{\lambda} \equiv w_1(\lambda)-w_2(\lambda)$, and $ C_i^2\vert_{\lambda} \equiv C^2\big(v_i(\lambda),w_i(\lambda)\big)$ with $i=1,2$. Expression \eqref{mainformula} is the working formula of this paper. In the following, we apply this formula in different collapse backgrounds, for three different families of diamonds. As we shall see, each one of them corresponds to a familiar notion of entropy discussed in the literature---now free of any UV ambiguity.


\section{Three notions of entropy for gravitational collapse}
\label{sec: gravitational collapse}

In this section we specialize the notion of entanglement entropy production to dynamical spacetimes representing gravitational collapse (and subsequent black hole evaporation). Considering various different types of causal domains, this leads to precise definitions of the notion of ``thermal entropy of Hawking quanta'', of Sorkin's ``exterior entropy'' \cite{Sorkin:2014kta} and of Page's ``radiation entropy'' \cite{Page:1993bv}. 

\begin{figure*}
\centering
\includegraphics[width=0.3\textwidth]{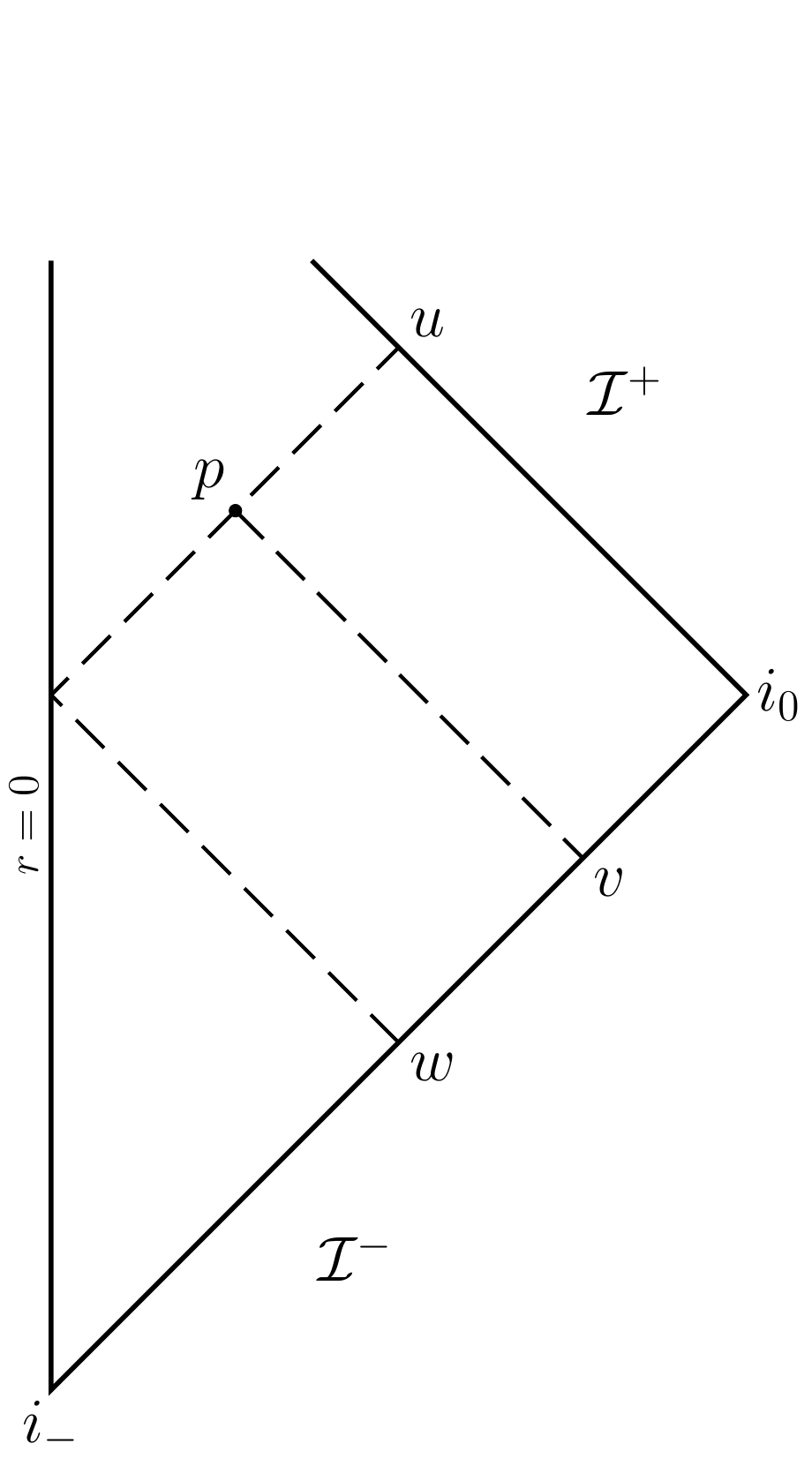}
\caption{Definition of the shadow coordinates $(v,w)$. Here $u$ and $v$ are affine coordinates on $\mathcal{I}^{+}$ (resp. $\mathcal{I}^{-}$), with $u=f(w)$.}
\label{fig:shadow}
\end{figure*}

\subsection{Collapse geometry and dimensional reduction}

Given a pair of double-null coordinates $(v,w)$ in the time-radius plane, the metric of a general spherically symmetric spacetime can be written as \cite{Roman:1983zza}
\begin{equation}
ds^2=- C^2(v,w)\,dv\,dw+r^2(v,w)\,(d\theta^2+\sin^2\theta\,d\phi^2)
\label{eq:roman}
\end{equation}
where $ C^{2}(v,w)$ is a conformal factor in the time-radius plane, $r(v,w)>0$ is the area radius, and $d\theta^2+\sin^2\theta\,d\phi^2$ is the metric on the unit $2$-sphere. In collapse settings where spacetime is asymptotically flat in the past, there is a natural choice for globally defined null coordinates $(v,w)$. Denote past null infinity $\mathcal{I}^{-}$, and pick an affine parameter along $\mathcal{I}^{-}$ in the time-radius direction. Given a point $p$, we define the \emph{shadow coordinates} $v(p)$ and $w(p)$ as the affine parameters of the two radial null rays which meet at $p$, with $w(p)$ the coordinate of the null ray bouncing at the centre.\footnote{There is of course a two-parameter family of such coordinates, following from the ambiguity of the affine parametrization of $\mathcal{I}^{-}$.} By construction, the shadow coordinates are such that $w\leq v$. Furthermore, the center $r(v,w)=0$ has equation $w=v$, and past (resp. future) null infinity corresponds to $w\to -\infty$ (resp. $v \to+\infty$); without loss of generality, we require that $\lim_{w\to-\infty} C^{2}(v,w)=1$. Note that, being defined using data at $\mathcal{I}^{-}$, $v$ and $w$ are well-defined also in the presence of a future event horizon. See Fig.~\ref{fig:shadow}.

Consider now a minimally coupled massless scalar field prepared in the Minkowski vacuum state at $\mathcal{I}^-$. At sufficiently high energy/frequency, the $s$-wave modes of the field are described by a $(1+1)$-dimensional field theory on the curved background $ds^2=- C^2(v,w)\,dv\,dw$ \cite{Birrell:1982ix,Fabbri:2005mw}. As is well known, this ``geometric optics'' approximation allows us to obtain the renormalized vacuum energy-momentum tensor $\langle T_{ab}\rangle$ in closed form \cite{Davies:1976ei}, and connects the physics of black hole evaporation with more intuitive settings, such as moving mirror systems \cite{Davies:1976hi,Davies:1977yv}. Thanks to the Holzhey-Larsen-Wilczek formula \cite{Holzhey:1994we}, this approximation also permits explicit computations of vacuum entanglement entropy in collapse spacetimes. We now turn to various implementations of this observation.

\subsection{Finite diamonds: entropy in a thick shell}

Our first example of vacuum entanglement entropy in collapse spacetimes is directly inspired by \cite{Holzhey:1994we}, where the authors discussed the ``geometric entropy'' of a finite segment in $(1+1)$ dimensions. In the context of spherically symmetric gravitational collapse, we can consider similarly the entanglement entropy in a thick spherical shell far from the centre $r=0$, as follows.

In asymptotically flat spacetimes, there exists a time coordinate $t^{*}$ such that $\partial_{t^{*}}$ is an asymptotic Killing vector at large radii $r$. Denote $r_{1}$ and $r_{2}$  two integral curves of $\partial_{t^{*}}$, and define $D_{t^{*}}$ as the domain of dependence of the segment $[r_{1},r_{2}]$ lying within the constant-$t^{*}$ surface. Given a reference time $t_{0}^{*}$, we can define the entanglement entropy production $\Delta S_{\textrm{shell}}(t^{*})$ in the thick spherical shell $[r_{1},r_{2}]$ at time $t^{*}$ as the excess entropy of $D_{t^{*}}$ with respect to $D_{t^{*}_{0}}$,
\be
\Delta S_{\textrm{shell}}(t^{*})\equiv S_{+}(D_{t^{*}})-S_{+}(D_{t^{*}_{0}}).
\ee
From (\ref{mainformula}), we have

\be\label{defholzhey}
\Delta S_{\textrm{shell}}(t^*)\equiv\f{1}{12}\log\f{\Delta v(t^*)^{2}\Delta w(t^*)^{2}\, C^{2}_{1}(t^*)\, C^{2}_{2}(t^*)}{\Delta v(t^*_{0})^{2}\Delta w(t^*_{0})^{2}\, C^{2}_{1}(t^*_{0})\, C^{2}_{2}(t^*_{0})}
\ee
with $C^{2}_{i}(t^*) \equiv C^{2}(v(t^*,r_{i}),w(t^*,r_{i}))$, $\Delta v(t^*)\equiv v(t^*,r_{2})-v(t^*,r_{1})$ and $\Delta w(t^*) \equiv w(t^*,r_{2})-w(t^*,r_{1})$. This quantity will be referred to as the \emph{shell entropy}.


\begin{figure*}
\centering
\includegraphics[width=0.28\textwidth]{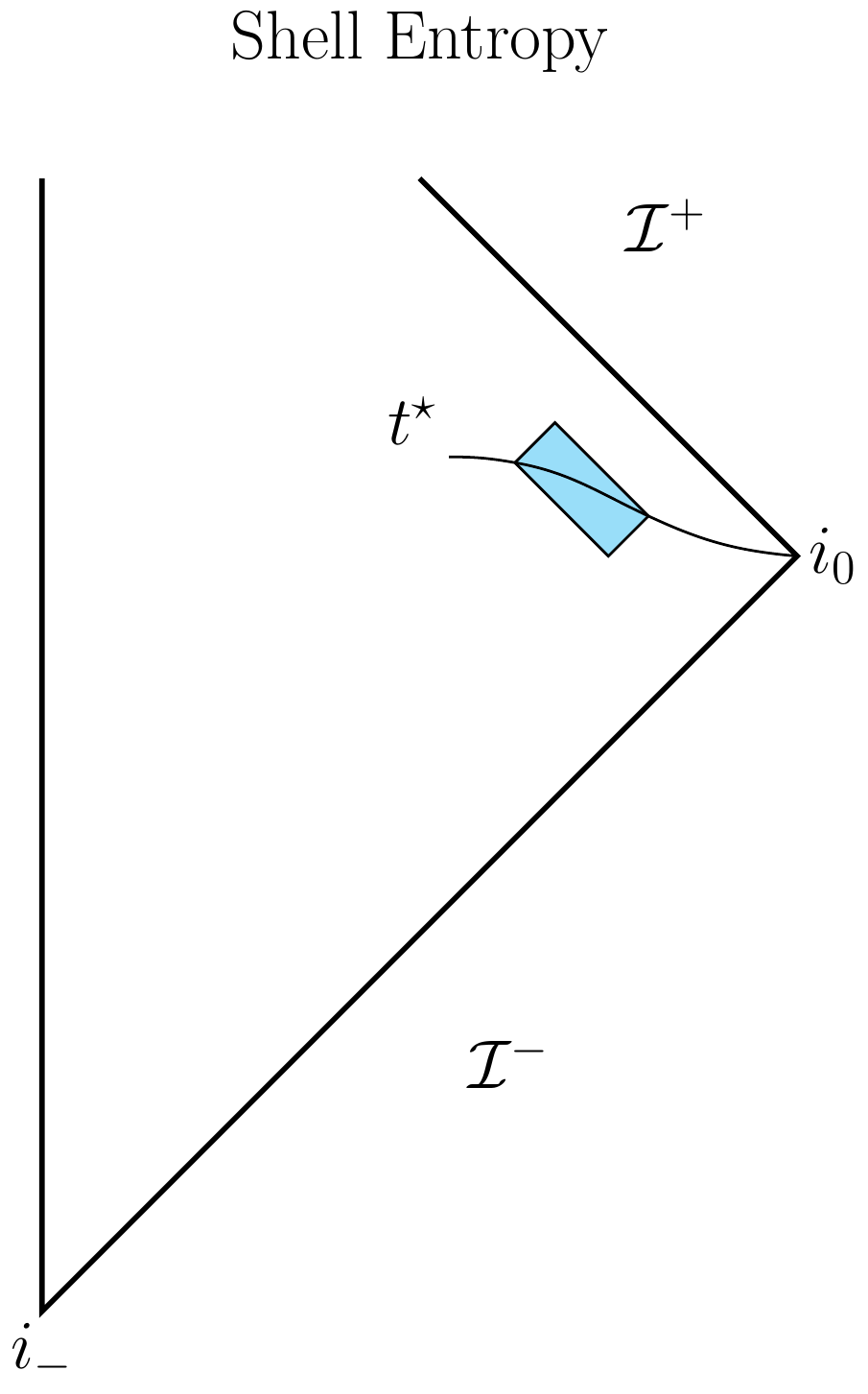}
\qquad
\includegraphics[width=0.28\textwidth]{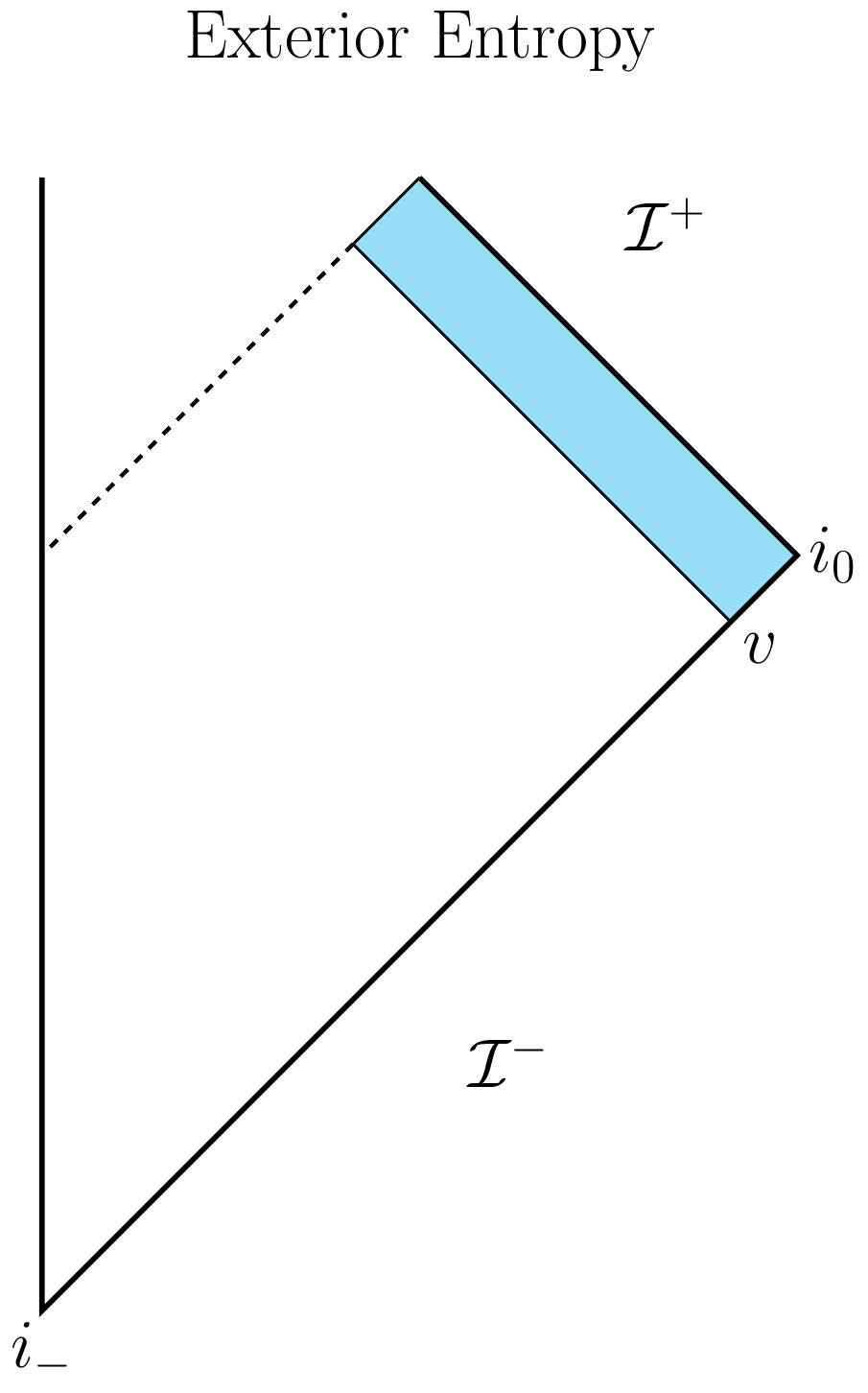}
\qquad
\includegraphics[width=0.28\textwidth]{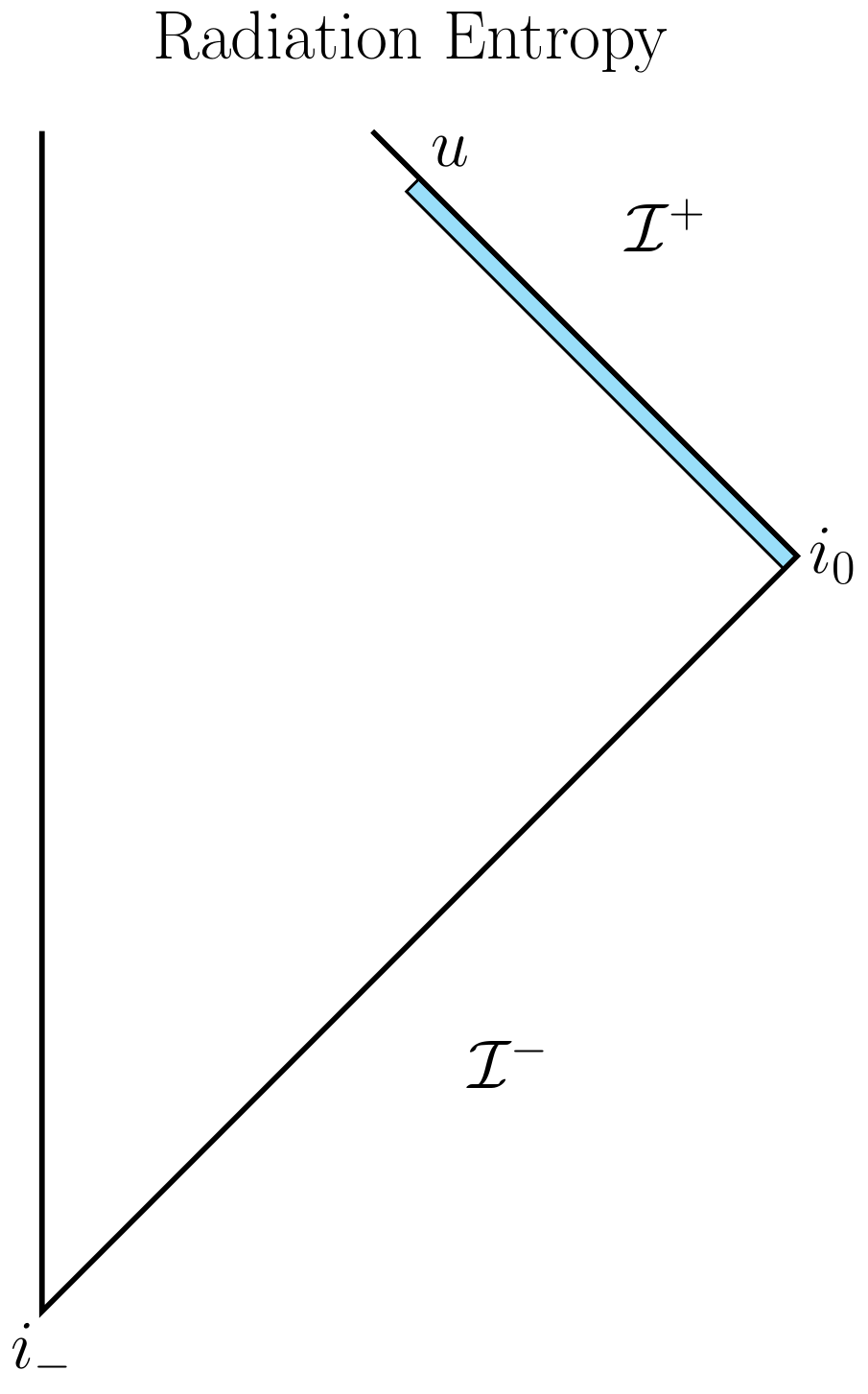}

\caption{Diamonds $D_{\lambda}$ involved in the definition of the shell, exterior and radiation entropies, with $\lambda=t^{*},v,u$ respectively.}
 \label{}
        \end{figure*}

\subsection{Infinite diamonds I: exterior entropy}
\label{sec: exterior entropy}

In \cite{Sorkin:2014kta,Bombelli:1986rw} Sorkin \emph{et al.} considered the entanglement entropy of quantum fields in the exterior of a black hole, a quantity that plays an important role in the generalized second law of thermodynamics \cite{Sorkin:1997ja,Wall:2009wm,Wall:2010cj,Wall:2011hj}. Clearly this quantity is UV divergent and once regularized it depends explicitly on the cut-off. Here we define the ``exterior entropy'' $\Delta S_{\textrm{ext}}$ of a black hole as the excess entanglement entropy of a causal domain with a corner at the horizon and a corner at spatial infinity $i_0$. This quantity is defined using the causal-splitting regularization and is manifestly independent of the cut-off. 

Consider a spacetime with a future event horizon $H$ with equation $w=w_{H}$. On a Cauchy slice $\Sigma$ that intersects $H$, the exterior of the black hole is the region $R=\Sigma\,\cap\, J^-(\mathcal{I}^+)$. In covariant terms this region corresponds to a causal domain $D_{p,q}$ with corners $p\in H$ and $q\to i_{0}$, where $i_{0}$ denotes spatial infinity. Using the shadow coordinate $v(p)$ as a parameter along $H$, we can define the \emph{exterior entropy} as the entanglement entropy production at `time' $v$
\begin{equation}\label{defsorkin}
\Delta S_{\textrm{ext}}(v)\equiv \lim_{q\to i_{0}}\Big(S_{+}(D_{p_{v},q})-S_{+}(D_{p_{0},q})\Big),
\end{equation}
where $p_{v}=(v,w_{H})$ and $p_{0}=(v_0,w_{H})$ is a reference point on $H$. Denoting $ C_{H}^{2}(v)\equiv  C^{2}(v,w_{H})$
the conformal factor at the horizon, the exterior entropy simplifies to
\be
\Delta S_{\textrm{ext}}(v)=\frac{1}{12}\log\frac{ C^2_H(v)}{ C^2_H(v_0)}.
\ee 
We will see in the next section how $\Delta S_{\textrm{ext}}$ relates to the Bekenstein-Hawking entropy of the black hole \cite{Bekenstein:1973ur,Hawking:1974sw}.

\subsection{Infinite diamonds II: radiation entropy}

In \cite{Page:1993bv} Page introduced the entanglement entropy of fields at future null infinity $\mathcal{I}^{+}$ as a measure of the ``age'' of an evaporating black hole. Based on an analogy with the entanglement entropy of finite-dimensional quantum systems, he argued that the entropy in the radiation must have two phases as a function of time: a growing phase, corresponding to the emission of thermal Hawking radiation, and a decreasing phase, corresponding to the ``purification'' of the Hawking quanta. 

From the perspective of this paper, the ``radiation entropy'' can be defined as the entanglement entropy production in a causal domain which approaches $\mathcal{I}^{+}$ asymptotically, as follows. Outgoing null geodesics that reach future null infinity provide a canonical map between $\mathcal{I}^-$ and $\mathcal{I}^+$. This map can be written as $u=u(w)$ where $u$ is an affine null coordinate at $\mathcal{I}^+$. We fix the ambiguity in $u$ by requiring that $(i)$ the null vectors $l=\partial_v$ and $n=\partial_u$ are canonically normalized at spatial infinity $i_0$, i.e. $l\cdot n\to-1$ there, and $u(0)=0$. Given this prescription, the  mapping $u=u(w)$ can be written  as
\be\label{uofw}
u(w)=\int_{0}^{w}  C^2_{\mathcal{I}^+}(w')\,dw'
\ee
where $C^2_{\mathcal{I}^+}(w)\equiv\lim_{v\to\infty} C^2(v,w)$ is the conformal factor at $\mathcal{I}^{+}$. Now, given a point $p_{u}\in \mathcal{I}^{+}$ with coordinate $u$, we define the \emph{radiation entropy} at retarded time $u$ as the limit
\be\label{defradiationentropy1}
\Delta S_{\textrm{rad}}(u)\equiv\lim_{p\to p_{u}}\,\lim_{p_{0}\to i_{0}}\,\lim_{r\to i_{0}}\Big(S_{+}(D_{p,r})-S_{+}(D_{p_{0},r})\Big).
\ee
This gives
\begin{equation}\label{defradiationentropy2}
\Delta S_{\textrm{rad}}(u)=\frac{1}{12}\log\, C^2_{\mathcal{I}^+}\!\big(w(u)\big).
\end{equation}
Equation \eqref{defradiationentropy1} can be rewritten as
\be\label{Sradraytracingmapping}
\Delta S_{\textrm{rad}}(u)=-\frac{1}{12}\log\,\dot{w}(u)
\ee
where the dot denotes derivatives with respect to $u$ and $w(u)$ is the inverse of the function (\ref{uofw}). Thus, radiation entropy is nothing but the \emph{logarithmic redshift} of outgoing rays. Equivalently, we can express $\Delta S_{\textrm{rad}}(u)$ in terms of the so-called \emph{peeling function}
\be
\kappa(u)\equiv-\f{\ddot{w}(u)}{\dot{w}(u)}
\ee
as
\be\label{Sradpeeling}
\Delta S_{\textrm{rad}}(u)=\f{1}{12}\int_{-\infty}^{u}\kappa(u')\,du',
\ee
as found in \cite{Bianchi:2014qua,Bianchi:2014vea}. Equation \eqref{Sradpeeling} provides us with an intuitive geometric interpretation of the ``Page curve'' $\Delta S_{\textrm{rad}}(u)$ often discussed in the black hole literature: $\Delta S_{\textrm{rad}}(u)$ grows when the separation between neighboring outgoing geodesics increases, i.e. they are \emph{peeled} ($\kappa(u)>0$), and decreases when their separation decreases, i.e. they are \emph{squeezed} ($\kappa(u)<0$). We emphasize that from \eqref{Sradraytracingmapping} and \eqref{Sradpeeling} and for a given spacetime, one can \emph{check} whether $\Delta S_{\textrm{rad}}(u)$ has the characteristic up-then-down behavior posited by Page \cite{Page:1993bv}. Finally, we note that 
\begin{equation}
  \lim_{u\to\infty}\Delta S_{\textrm{rad}}(u)=0
\end{equation}
is neither a necessary nor a sufficient condition for unitarity of quantum evolution of a massless field from $\scri^-$ to $\scri^+$. However, as will be illustrated below, \emph{finiteness} of $\Delta S_{\textrm{rad}}(u)$ at all times is necessary.

\section{Vacuum thermalization in a Vaidya spacetime}
 \label{sec: Vaidya} 

In this section we consider the simplest possible model of gravitational collapse: the two-dimensional Vaidya ingoing shell. Although too simple to address the important issue of unitarity in black hole evaporation, this model illuminates the nature of entanglement in the Hawking thermalization process and provides the basic intuition underlying the notions of shell, exterior and radiation entropy.

\subsection{Geometry of the Vaidya collapse}
\label{sec: geodesics}

In advanced Eddington-Finkelstein coordinates $(v,r)$, the spacetime metric for a Vaidya ingoing shell at $v=v_s$ reads
\be\label{vaidya}
ds^{2}=-\left(1-\f{2M}{r}\Theta\big(v-v_s\big)\right)dv^{2}+2dvdr,
\ee
where $\Theta$ is the Heaviside function and $M$ the mass of the incipient black hole. This spacetime consists of a flat patch in the past of the shell ($v<v_s)$, and of a Schwarzschild patch in the future ($v>v_s$). Finally, we set $v=0$ to be the ray that crosses the infalling shell at $r=3M$. 

The expression of the Vaidya metric \eqref{vaidya} in double null coordinate $(v,w)$ is easily obtained. The coordinate $v$ labels infalling null geodesics. An infalling null geodesic with advanced time $v=w$ is reflected at the centre $r=0$ and results in an outgoing null geodesic $r(v,w)$ with $v\geq w$. We use the coordinate $w$ to label outgoing null geodesic. The trajectory $r(v,w)$ is obtained by solving the null geodesic equation
\be
\f{dr}{dv}=\f{1}{2}\left(1-\f{2M}{r}\Theta\big(v-v_s\big)\right).
\ee
with initial condition $r(w,w)=0$. We obtain
\begin{numcases}{r(v,w)=}
\frac{v-w}{2} & for $v < v_s$, \nonumber\\[.5em]
\textstyle 2M\left\{1+W\Big(\big(\frac{v_s-w}{4M}-1\big)\exp\left[\frac{v-w}{4M}-1\right]\Big)\right\}
& for $v \geq v_s$,\label{rBH}
\end{numcases}
where $W$ is the Lambert function.\footnote{The Lambert $W$-function is defined for $x\geq -e^{-1}$ as the unique solution of the equation $W(x)e^{W(x)}=x$. It satisfies $W(x)\sim x$ as $x\to0$, $W(x)\to \infty$ as $x\to \infty$, and $W'(x)=W(x)/[x(1+W(x))]$.} The conformal factor in (\ref{eq:roman}) in shadow coordinates $(v,w)$ is then obtained computing $ C^{2}(v,w)=-2\pp_{w}r(v,w)$, 
\begin{numcases}{ C^{2}(v,w)=}
\qquad 1 & for $v \leq v_s$, \nonumber\\[1em]
\frac{w-v_s}{w-v_s+4M}\,\frac{W\Big(\big(\frac{v_s-w}{4M}-1\big)\exp[\frac{v-w}{4M}-1]\Big)}{1+W\Big(\big(\frac{v_s-w}{4M}-1\big)\exp[\frac{v-w}{4M}-1]\Big)} & for $v \geq v_s$\label{omegaBH}.
\end{numcases}
Using formula \eqref{uofw} and the value of the conformal factor at $\mathcal{I}^+$,
\be\label{omegapageV}
 C^2_{\mathcal{I}^+}(w)=\lim_{v\to +\infty} C^{2}(v,w)=\frac{w-v_s}{w-v_s+4M}\,,
\ee
we find
\be\label{uwV}
u(w)=w-4M\,\log\frac{v_s-4M-w}{v_{s}-4M}
\ee

\begin{figure*}
\centering
\includegraphics[width=0.3\textwidth]{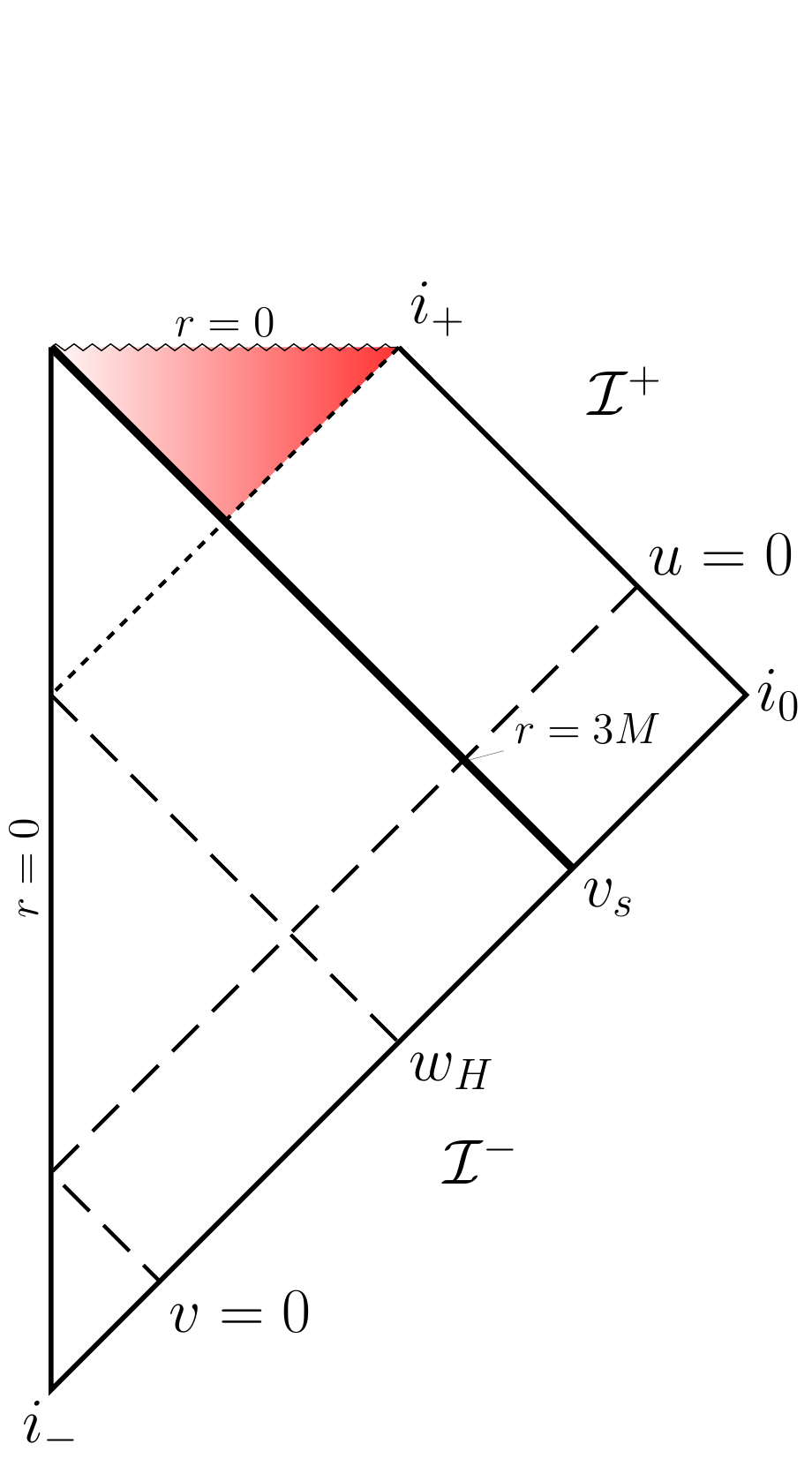}
\hfill
\includegraphics[width=0.55\textwidth]{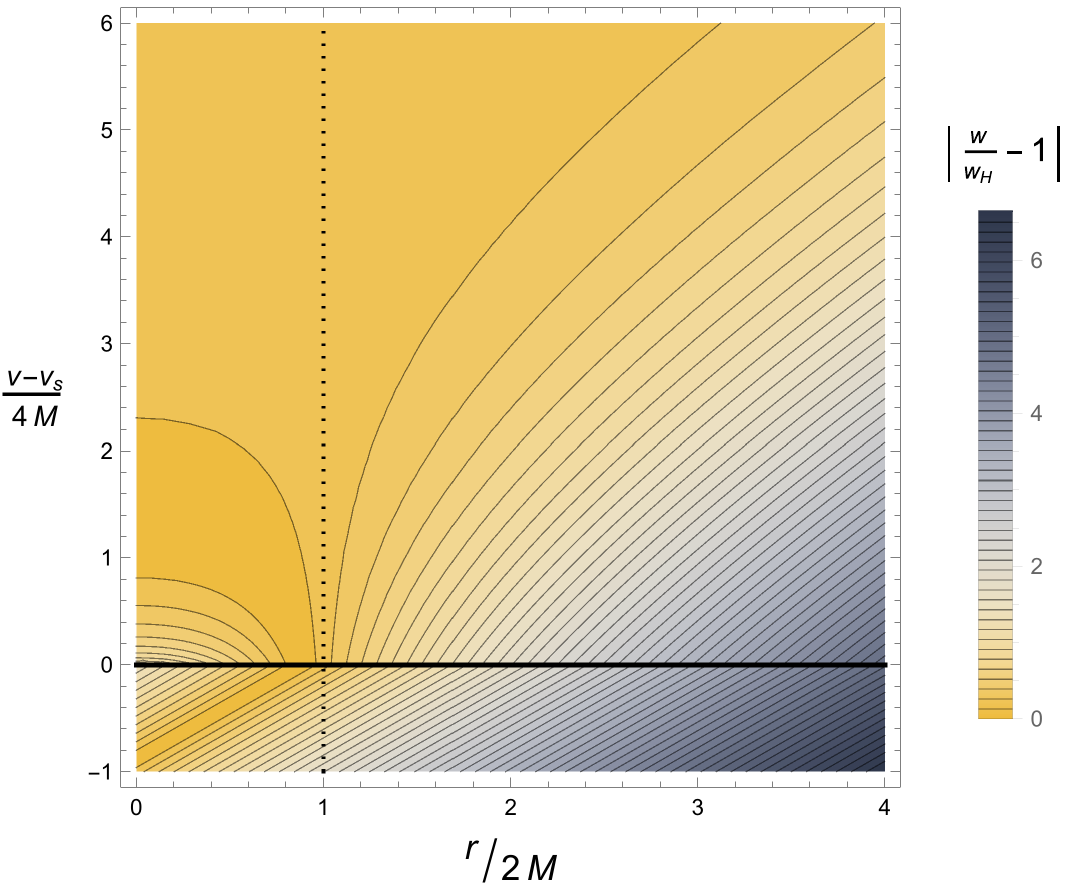}
\caption{Geometry of the Vaidya collapse model. Left: Carter-Penrose diagram, with the trapped region shaded. Right: outgoing null geodesics in Eddington-Finkelstein coordinates. Both diagrams show the infalling shell (thick line) and the event horizon (dotted line).}
\label{fig:geo_vaidya}
\end{figure*}

The map $u=u(w)$ is defined only for $w\leq v_s-4M$ which identifies the presence of an event horizon $H$ at
\begin{equation}
w_H\equiv v_s-4M\,.
\end{equation}
The Carter-Penrose diagram for the metric (\ref{vaidya}) and the outgoing null geodesics are in Fig.~\ref{fig:geo_vaidya}.

Equations \eqref{rBH}, \eqref{omegaBH} and \eqref{uwV} contain the geometric information required for the evaluation of the shell, exterior and radiation entropies in the Vaidya model of gravitational collapse.  


\subsection{Shell entropy and the thermodynamics of Hawking quanta}
\label{eq:shell-entropy}

\begin{figure}
\includegraphics[scale=.8]{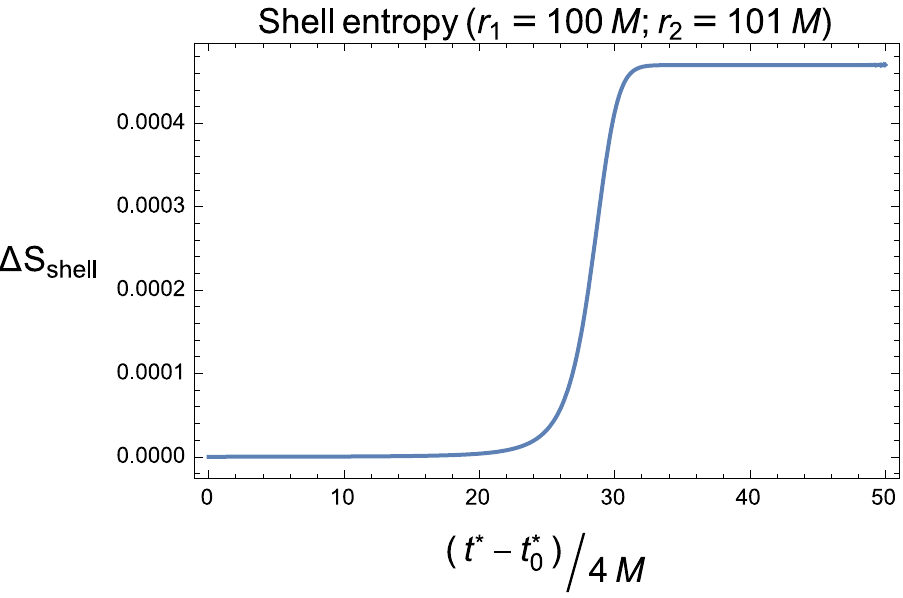}\hfill
\includegraphics[scale=.8]{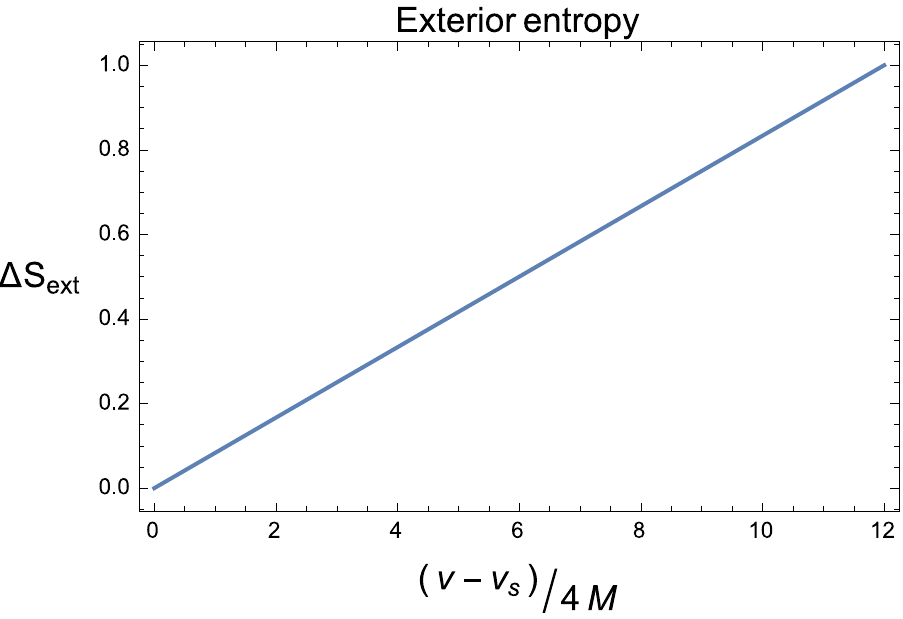}\hfill
\caption{Entanglement entropy production in a Vaidya spacetime with $M=10m_P$. Left: Entropy in a spherical shell $100M\leq r\leq 101M$, converging to the thermal value \eqref{Stherm}. Right: Entropy of the black hole exterior, linearly increasing in time.}
\label{plotsV}
\end{figure}

Consider a stationary spherical shell $[r_{1},r_{2}]$ far away from the incipient black hole, $r_2>r_{1}\gg 2M$, and denote $t^{*}\equiv v-r$ the Finkelstein time coordinate ($\pp_{t^{*}}$ is an asymptotic Killing vector at $r\rightarrow\infty$). The double null coordinates $p_i=\big(v_i(t^*),w_i(t^*)\big)$, $i=1,2$, of the two walls of the spherical shell at time $t^*$ are easily determined using $v_i(t^*)=r_i+t^*$ and $r\big(v_i(t^*),w_i(t^*)\big)=r_i$.
Using formula \eqref{defholzhey} and \eqref{omegaBH} we find that the excess entanglement entropy of the shell at time $t^*$ with respect to the reference time $t^{*}_{0}\equiv v_s-2M$ is
\begin{align}
\Delta S_{\textrm{shell}}(t^{*})=&\frac{1}{12} \log\Bigg[\frac{\Big(W\big(q_1 \exp[-(t^*-t^*_0)/4M]\big)-W\big(q_2 \exp[-(t^*-t^*_0)/4M]\big)\Big)^2}{\Big(W(q_1)-W(q_2)\Big)^2}\times\nonumber\\[.5em]
 &  \;\;\times\frac{\Big(1+W\big(q_1 \exp[-(t^*-t^*_0)/4M]\big)\Big)\Big(1+W\big(q_2 \exp[-(t^*-t^*_0)/4M]\big)\Big)}{W\big(q_1 \exp[-(t^*-t^*_0)/4M]\big)W\big(q_2 \exp[-(t^*-t^*_0)/4M]\big)}\Bigg] \,,
\label{holzheyV}
\end{align}
where $q_i\equiv \frac{r_i}{2M}\exp[\frac{r_i+2M}{4M}]$. This function of Finkelstein time $t^{*}$ is plotted in Fig.~\ref{plotsV}. 
Using the properties of the Lambert $W$ function, one can show from this expression that the shell entropy $\Delta S_{\textrm{shell}}(t^{*})$ starts at zero and, after a transient starting at $t^{*}=\mathcal{O}(r_{1})$ and lasting a time $\Delta t^{*}=\mathcal{O}(M+\Delta r)$, it reaches a plateau where it converges to the final value
\be\label{boxentropy}
\Delta S_{\textrm{shell}}(\infty)=\f{1}{6}\log \left(\f{\sinh[\Delta r/8M]}{\Delta r/8M}\right).
\ee
Here $\Delta r\equiv r_{2}-r_{1}$. As noted by Holzhey \emph{et al.} \cite{Holzhey:1994we}, this value coincides with the entropy of a thermal state. More precisely, defining  $S_{\textrm{therm}}(T)$ as the entanglement entropy of a mixed thermal state at temperature $T$ for an interval $\Delta r=r_2-r_1$ in flat Minkowski space, one finds that the excess entanglement entropy is
\begin{equation}
S_{\textrm{therm}}(T)-S_{\textrm{therm}}(0)\,=\,\frac{1}{6}\log \left(\frac{\sinh [\pi\, \Delta r\, T]}{\pi\, \Delta r\, T} \right)\,,
\label{Stherm}
\end{equation}
where the entanglement entropy of the vacuum $S_{\textrm{therm}}(T=0)$ has been subtracted \cite{Calabrese:2004eu}.  Comparing \eqref{Stherm} to \eqref{boxentropy} we see that excess entanglement entropy in a spherical shell captures the thermal nature of the Hawking radiation at temperature $T_{H}\equiv(8\pi M)^{-1}$.

Note that, in the limit of a thick shell, $\Delta r\equiv r_2-r_1\gg 2M$, this excess entanglement entropy reduces to
\begin{equation}
\Delta S_{\textrm{shell}}(\infty)\simeq\frac{\Delta r}{48M},
\label{largeboxlimit}
\end{equation}
i.e. the excess entanglement entropy of a thick shell surrounding the black hole is \emph{extensive}. An extensive entropy is the typical behavior of a thermal system. Indeed, the thermal entropy of a gas of massless scalars at temperature $T$ in a one-dimensional box of size $\ell$, as computed from standard statistical mechanics, is $S_{\textrm{therm}}(T)=\pi T\Delta r/6$. At the Hawking temperature $T=(8\pi M)^{-1}$, this matches precisely the result \eqref{largeboxlimit}. Therefore the excess entanglement entropy $\Delta S_{\textrm{shell}}(\infty)$ describes the thermal entropy of the Hawking radiation.

In the case of a \emph{thin} shell ($\Delta r\ll 2M$), on the other hand, formula (\ref{boxentropy}) results in an excess entanglement entropy that is sub-extensive
\begin{equation}
\Delta S_{\textrm{shell}}(\infty) \simeq\left(\frac{\Delta r}{48M}\right)^2\,,
\label{eq:DSi}
\end{equation}
and smaller than the thermal entropy $S_{\text{therm}}=\pi T_{H}\Delta r/6$ at the Hawking temperature. This phenomenon can be understood as follows. If a box is smaller than the typical wavelength $\lambda\sim T^{-1}$ of the thermal radiation, the Planckian distribution is cut off at the size of the box and the entropy is not captured by standard statistical mechanics---the excess entanglement entropy provides a finer description of the entropy of the system. This finer description coincides with the one obtained for the entanglement entropy of a thermal state in Minkowski space. 

\subsection{Exterior entropy and the generalized second law}

Next we consider the entropy of the black hole exterior defined in Sec.~\ref{sec: exterior entropy}. From \eqref{omegaBH} the conformal factor at a point $v$ on the event horizon $H=\{(v,w)|w=w_H\equiv v_s-4M\}$ is
\be
 C^{2}_{H}(v)= C^{2}(v,w_H)= \exp\left[\,\frac{v-v_s}{4M}\right],
\ee 
hence, taking as reference value for the entropy $v_0=v_s$, we get 
\be\label{sorkinV}
\Delta S_{\textrm{ext}}(v)=\f{1}{12}\log \frac{ C^{2}_{H}(v)}{ C^{2}_{H}(v_s)} =\f{v-v_s}{48M}.
\ee
Thus, from the perspective of the exterior, the black produces entanglement entropy at the constant rate $1/48M$. This is consistent with thermodynamical expectations \cite{Zurek:1982vq}. According to conventional thermodynamics, the entropy radiated by a blackbody at temperature $T$ in empty space (in $d$ spatial dimensions) satisfies 
\be\label{thermo}
T\,\Delta S_{\text{therm}}= \Delta E+p\, \Delta V=\f{d+1}{d} \Delta E,
\ee
where $\Delta E$ is the energy radiated and $p\, \Delta V$ the work term due to the radiation pressure. According to \eqref{thermo}, a two-dimensional blackbody at temperature $T_{H}$ emitting an energy flux $F_{H}=\pi T_{H}^{2}/12$---as is the case for a black hole---should emit an entropy
\be
\Delta S_{\text{therm}}=2\,\f{F_{H}}{T_{H}}\,\Delta v=\f{\Delta v}{48M}.
\ee
This is precisely what we found for the exterior entropy \eqref{sorkinV}. 

Another interesting observation about \eqref{sorkinV} regards the relationship between the exterior entropy and the Bekenstein-Hawking entropy $S_{\textrm{BH}}\equiv A_{H}/{4\ell_P^2}$, where $A_{H} \equiv 16\pi M^{2}$ is the area of the event horizon \cite{Bekenstein:1973ur,Hawking:1974sw}. By definition of $S_{\textrm{BH}}$, the quantity $T_{H}\Delta S_{\textrm{BH}}$ is equal to minus the radiated energy $\Delta E=F_{H}\Delta v$. Given our result that $\Delta S_{\textrm{ext}}=2\,\Delta E$, we have
\be
\Delta S_{\textrm{ext}}=-2\,\Delta S_{\textrm{BH}}.
\ee
This identity is consistent with the generalized second law of thermodynamics \cite{Bekenstein:1973ur,Hawking:1974sw}, according to which the Bekenstein-Hawking entropy of the black hole plus the entropy of exterior matter can never decrease,
\be\label{GSL}
\Delta S_{\textrm{matter}}+\Delta S_{\textrm{BH}}\geq0,
\ee 
if $\Delta S_{\textrm{matter}}$ interpreted as the exterior entropy production $\Delta S_{\textrm{ext}}$. From this perspective, the fact that 
\be
\Delta S_{\textrm{matter}}+\Delta S_{\textrm{BH}}=\f{\Delta v}{96M}>0
\ee 
can be interpreted as expressing the irreversibility of the Hawking evaporation process.

\subsection{Radiation entropy and a monotonic Page curve}
\label{sec:vaidya-Page}

\begin{figure}
\begin{center}
\includegraphics[scale=.78]{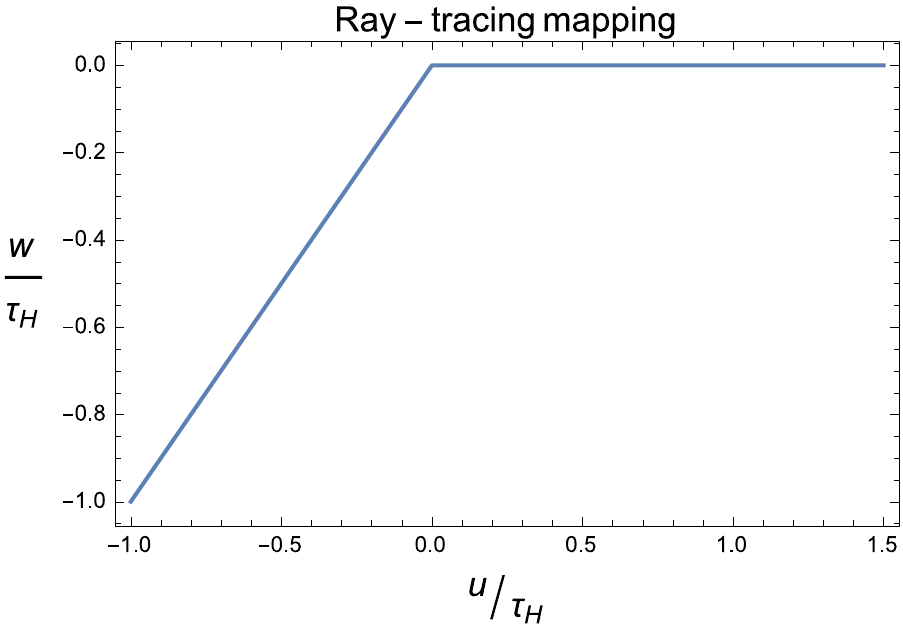}\hfill
\includegraphics[scale=.85]{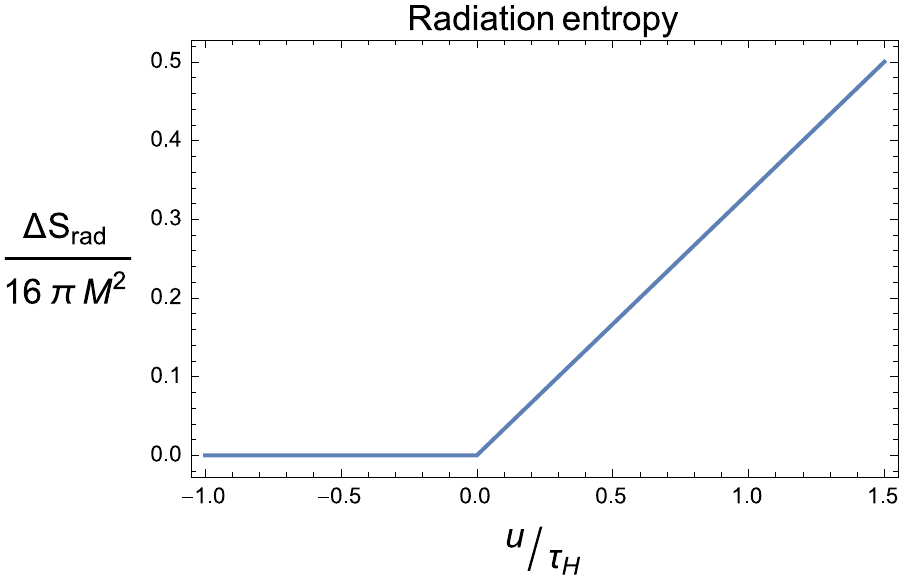}
\end{center}
\caption{Ray-tracing mapping $w=w(u)$ (left) and radiation entropy $\Delta S_{\text{rad}}(u)$ (right) in a Vaidya spacetime with $M=10m_P$.}
\label{fig:entropy_vaidya}
\end{figure}

Finally we compute the radiation entropy at $\mathcal{I}^{+}$. Inverting formula \eqref{uwV} for the map $u=u(w)$ between  $\mathcal{I}^{-}$ and  $\mathcal{I}^{+}$ we find
\be\label{uV}
w(u)=v_s-4M\left\{1+W\left(\frac{v_{s}-4M}{4M}\exp\left[-\frac{u-v_{s}+4M}{4M}\right]\right)\right\}.
\ee
Plugging \eqref{uV} into the expression \eqref{Sradraytracingmapping} for the radiation entropy, we arrive at 
\be
\Delta S_{\textrm{rad}}(u)=\f{1}{12}\log \left(\f{1+W\left(\frac{v_{s}-4M}{4M}\exp\left[-\frac{u-v_{s}+4M}{4M}\right]\right)}{W\left(\frac{v_{s}-4M}{4M}\exp\left[-\frac{u-v_{s}+4M}{4M}\right]\right)}\right).
\label{eq:entropy-vaidya}
\ee
In the limit $u\to -\infty$ the radiation entropy goes to zero, $\Delta S_{\textrm{rad}}(u)\to0$. At $u\approx 0$ (the retarded time when the in-falling shell reaches radius  $3M$), the entropy start growing monotonically, and at late times $u\to +\infty$, we have
\be
\Delta S_{\textrm{rad}}(u)\sim \f{u}{48M} .
\label{S-linear}
\ee
That is, the radiation entropy of a Vaidya black hole grows linearly and without bounds, corresponding to the monotonic ``Page curve'' shown in Fig.~\ref{fig:entropy_vaidya}. Such linear growth is again consistent with the Vaidya black hole acting---from the perspective of asymptotic observers at $\mathcal{I}^{+}$---as a steady source of thermal radiation.


\section{Radiation entropy: models of evaporation with and without horizon}
\label{sec: radiation}

In this section we extend our analysis of entanglement entropy production in gravitational collapse to other black-hole-like geometries: a collapsing star which stops just before crossing its Schwarzschild radius, a ``Hawking-like'' evaporating black hole (with event horizon), a nonsingular ``Bardeen-like'' evaporating black hole (without event horizon), and a model of black-to-white hole tunneling. We focus our attention on the radiation entropy $\Delta S_{\text{rad}}(u)$ measured at future null infinity and discuss the features of the corresponding ``Page curve''.

\subsection{Collapse to an $\varepsilon$-star}
\label{sec:eps-star}
\begin{figure*}
\centering
\includegraphics[width=0.25\textwidth]{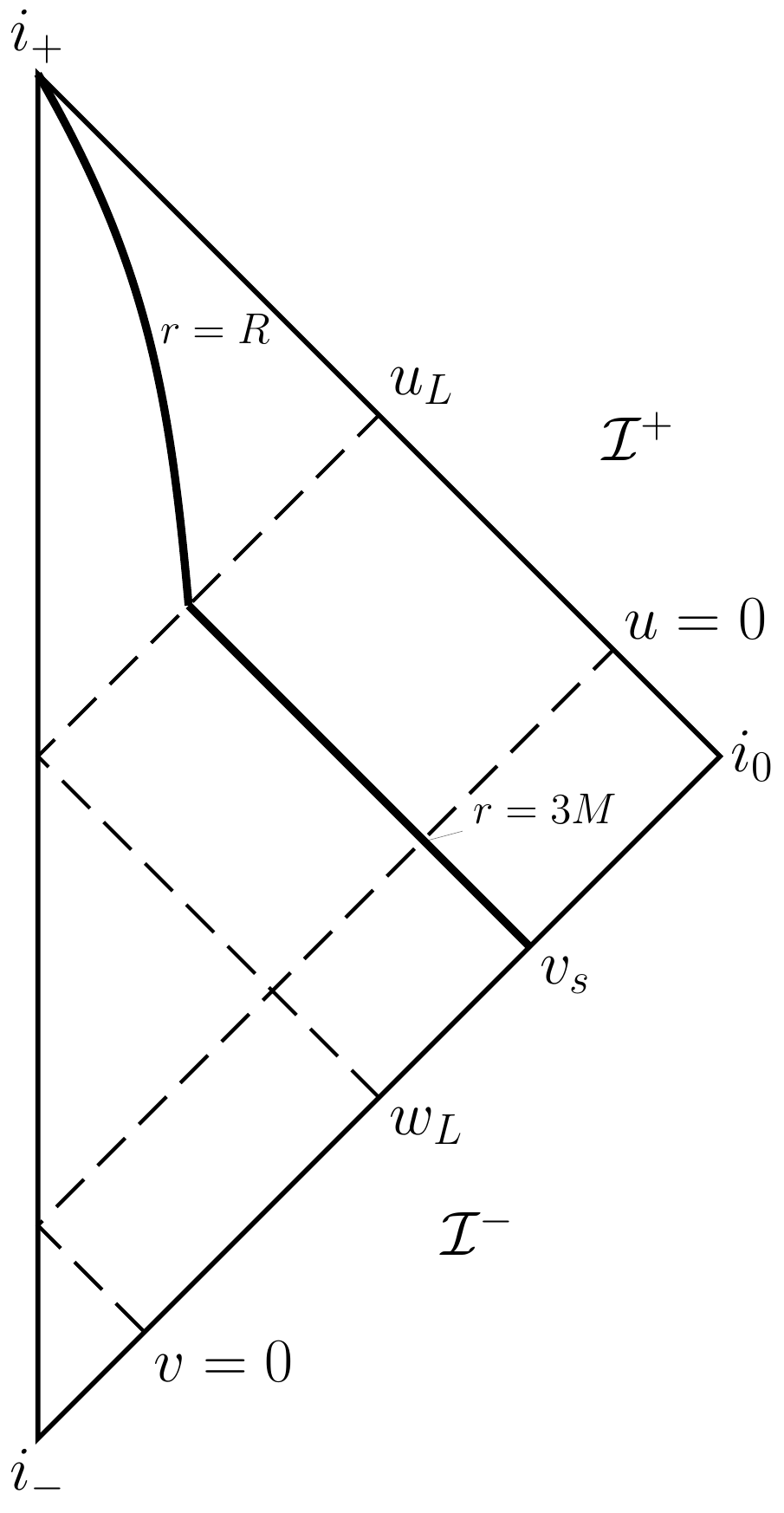}
\hfill
\includegraphics[width=0.55\textwidth]{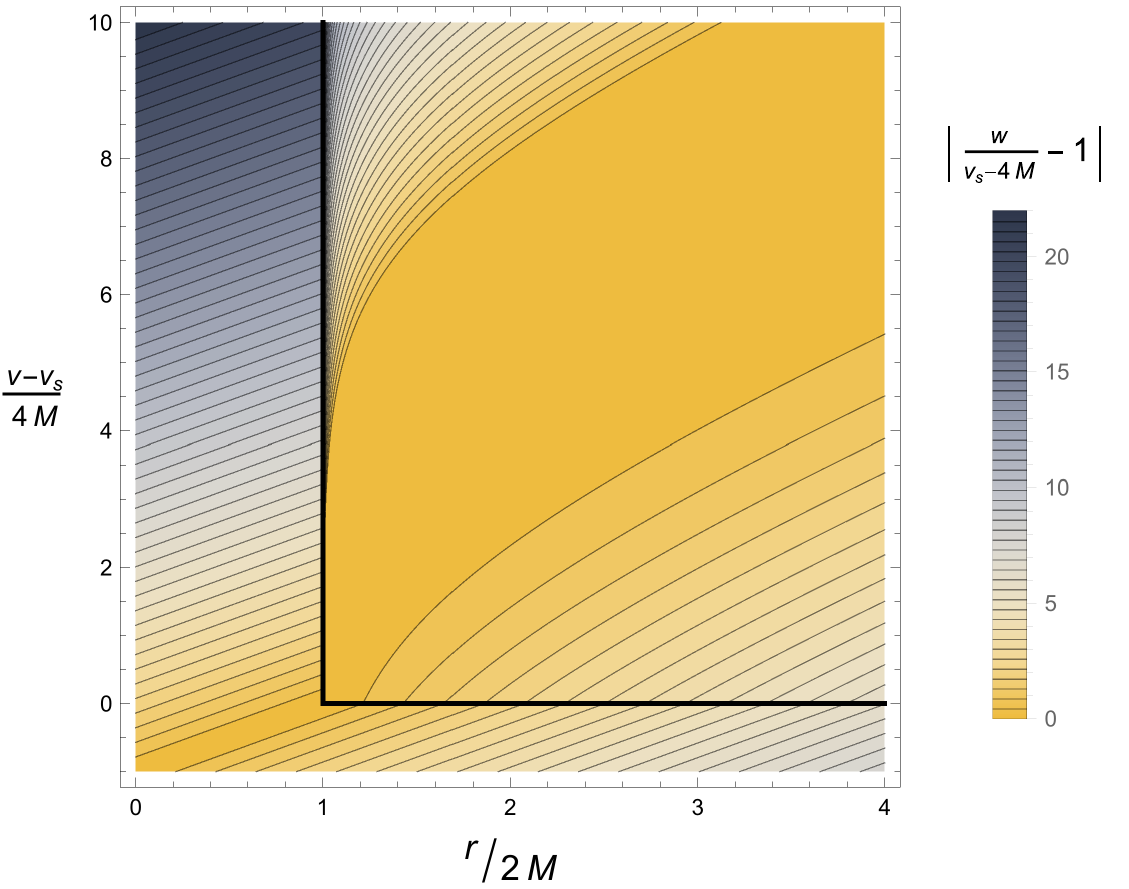}
\caption{Geometry of a `$\varepsilon$-star' collapse model, Eq. (\ref{quasi_BH}). Left: Carter-Penrose diagram. Right: outgoing null geodesics in Eddington-Finkelstein coordinates. Both diagrams show the infalling shell (thick  line). Note the existence of a ``Hawking region" ($w\sim v_s-4M$), where a thermal flux is recorded by stationary observers, in spite of the absence of a trapped region or event horizon.}
\label{fig:geo_quasi_BH}
\end{figure*}

\begin{figure}
\begin{center}
\includegraphics[scale=.77]{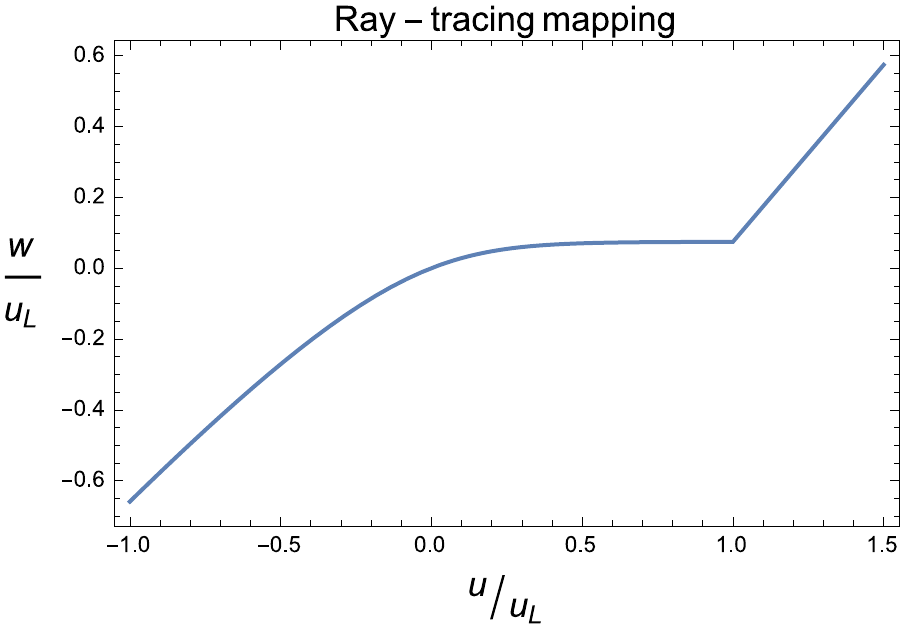}\hfill
\includegraphics[scale=.87]{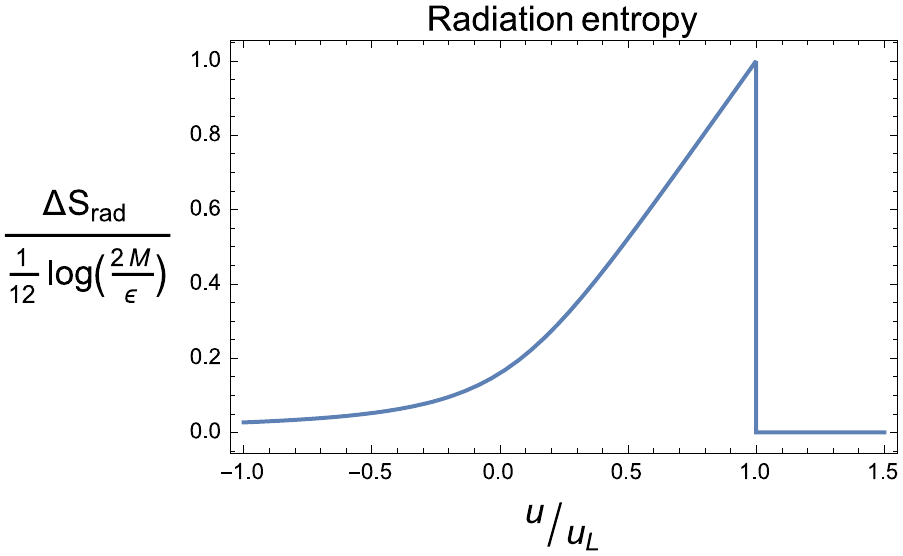}
\end{center}
\caption{Ray-tracing mapping $w=w(u)$ (left) and radiation entropy $\Delta S_{\text{rad}}(u)$ (right) in an `$\varepsilon$-star' collapse model with $M=10m_P$ and $\varepsilon=2M/10^3$. The duration of the thermal ``Hawking phase", before the sudden purification of the ougoing radiation, is $u_L\sim 4M \log(2M/\varepsilon)$.}
\label{fig:entropy_eps}
\end{figure}

Let us begin by repeating the above calculations in an model of gravitational collapse which leads to the formation a compact star, with no event or trapping horizon. In this model, considered in \cite{Stephens:1999dx,Paranjape:2013vi} and called $\varepsilon$-star in this paper, a collapsing shell of mass $M$ ``freezes'' at 
\begin{equation}
R\equiv 2M+\varepsilon\,,
\end{equation}
with $\varepsilon\ll 2M$, see Fig. \ref{fig:geo_quasi_BH}. The corresponding metric, with parameters $M$ and $\varepsilon$, is 
\be\label{quasi_BH}
ds^{2}=-\left(1-\f{2M}{r}\Theta\big(v-v_s\big)\Theta\big(R-r)\right)dv^{2}+2dvdr\,.
\ee
Solving the geodesic equation for outgoing light rays as in Sec.~\ref{sec: geodesics}, we find the canonical map $u\mapsto w(u)$ between $\mathcal{I}^+$ and $\mathcal{I}^-$. Defining $w_L$ as the advanced time when the in-falling shell stops, i.e. $r(v_s,w_L)=R$, the canonical map $w=w(u)$ is given by
\begin{numcases}{\!\!\!\!\!\!\!w(u)=}
v_s-4M\left\{1+W\left(\frac{v_{s}-4M}{4M}\exp\left[-\frac{u-v_{s}+4M}{4M}\right]\right)\right\} & if $u\leq u_{L}$\\[.5em]
u-u_{L}+v_s-4M\left\{1+W\left(\frac{v_{s}-4M}{4M}\exp\left[-\frac{u_{L}-v_{s}+4M}{4M}\right]\right)\right\}\!\! & if $u> u_{L}\,$\nonumber
\end{numcases}
where 
\be
u_{L}=v_s-4M-2\varepsilon+4M\log\left(\frac{v_s-4M}{2\varepsilon}\right)\sim 4M\log(2M/\varepsilon)
\label{ueps}
\ee 
is defined by $w(u_{L})=w_{L}$. The map $w(u)$ is onto: there is no event horizon and the spacetime with metric \eqref{quasi_BH} has the same causal structure as Minkowski space. Plugging \eqref{ueps} into the expression \eqref{Sradraytracingmapping} for the radiation entropy we find the ``Page curve'' shown in Fig.~\ref{fig:entropy_eps}. For $u<u_L$ the function $w(u)$ coincides with the one of the Vaidya spacetime \eqref{uV} and therefore the entropy of the radiation coincides with formula \eqref{eq:entropy-vaidya}. However this emission phase stops at the finite time $u_L$ where the radiation entropy reaches its maximum
\begin{equation}
\Delta S_{\text{rad}}^{\text{max}}=\frac{1}{12}\log\Big(1+2M/\varepsilon\Big)\sim \frac{1}{12}\log(2M/\varepsilon)\,.
\label{eq:Sradmax}
\end{equation}
The radiation emitted is approximately thermal with temperature $T=1/(8\pi M)$ for a finite time of order\footnote{Notice that a similar-looking expression has been discussed recently under the names ``information retention time'' \cite{Hayden:2007bc} and ``scrambling time'' \cite{Sekino:2008im,Susskind:2011wd}. Here it appears as the time during which a collapsing matter distribution that does not cross its Schwarzschild horizon cannot be distinguished from an incipient black hole.
} $4M\, \log(2M/\varepsilon)$. If a shell of one solar mass stops at a Planck length from the would-be event horizon, i.e. $\varepsilon=\ell_P$, the emission phase would last only $\Delta u \approx 10^{-3} \,\text{s}$. After this phase the entanglement entropy and the energy flux vanish as in Minkowski space. No information is lost.


\subsection{Singular black hole evaporation}
\label{sec:vaidya-shell-antishell}

In section \ref{sec:vaidya-Page} we found that the formation of a black hole by a collapsing shell produces radiation with a linearly increasing entanglement entropy at late times. The analysis however does not take into account the fact that, by energy conservation, the black hole loses mass during the emission process. Given the asymptotic flux $F_H=1/(768\pi M^2)$, one finds that in a finite time $\tau_H$ the mass of the black hole decreases to zero. Solving the equation $\dot{M}=-F_H$ one finds the evaporation time to be
\begin{equation}
\label{eq:Hawking-time}
\tau_H \simeq  \alpha\, \frac{M^3}{m_P^2}
\end{equation}
with $\alpha=256\pi$. We now consider a toy model of gravitational collapse and subsequent evaporation that can be solved analytically. In this model a black hole forms by the collapse of a null shell falling at the advanced time $v_s$. The mass of the black hole remains constant and equal to $M$ for a time $\sim\tau_H$ after the onset of Hawking evaporation, and then instantaneously vanishes. The evaporation process is modeled following \cite{Hiscock:1980ze,Hiscock:1981xb}. At the advanced time $\overline{v}_s$ and at distance $r=\overline{R}_s>2M$ (e.g. $\overline{R}_s=3M$) two null shells are produced: an out-going shell of mass $M$ and an in-going shell of mass $-M$. The out-going shell models the back-reaction on the metric of the positive energy flux brought by the Hawking radiation, the in-going shell models the mass loss of the black hole.

Let us define the function $r_0(v,w)$ as in (\ref{rBH}),
\begin{equation}
r_0(v,w)=2M\left\{1+W \left(\frac{v_s-4M-w}{4M}\exp\left[\frac{v-w}{4M}-1\right]\right)\right\},
\end{equation}
We call $(\overline{v}_s,\overline{w}_s)$ the point where the two null shells modeling the evaporation process are produced. Once fixed the advanced time $\overline{v}_s$, the condition that the production happens at the distance $r=\overline{R}_s$ determines $\overline{w}_s$ via the equation
\begin{equation}
r_0(\overline{v}_s,\overline{w}_s)=\overline{R}_s,
\label{eq:Hiscock}
\end{equation}
i.e. 
\begin{equation}
\overline{w}_s=v_s-4M\left\{1+W\left(\frac{\overline{R}_s-2M}{2M}\exp\left[\frac{2\overline{R}_s-4M-\Delta v}{4M}\right]\right)\right\}.
\end{equation}
where $\Delta v\equiv \overline{v}_s-v_s$.
The metric defining this model, with parameters $M$, $\overline{R}_s$ and $\Delta v$, has the form
\begin{equation}
ds^2=-F(v,r)\,dv^2+2\,dvdr
\label{eq:F(v,r)}
\end{equation}
with
\begin{numcases}{F(v,r)=}
\label{eq:hiscock-evap}
1 & if $v < v_s$, \\[.5em]
1-\frac{2M}{r}& if $v_s\leq v < \overline{v}_s$, \nonumber\\[.5em]
1-\frac{2M}{r}& if $v\geq \overline{v}_s$ and $r>r_0(v,\overline{w}_s)$, \nonumber\\[.5em]
1& if $v\geq \overline{v}_s$ and $r\leq r_0(v,\overline{w}_s)$. \nonumber
\end{numcases}
%
For $v<\overline{v}_s$ this metric coincides with the model of collapse discussed in section \ref{sec: Vaidya}. After the advanced time $\overline{v}_s$, however, the metric is flat for $r<r_0(v,\overline{w}_s)$ modeling the disappearance of the black hole. This geometry has a spacelike singularity at $r=0$ and a trapping horizon \emph{TH} at
\begin{equation}
w_{TH}\equiv v_s-4M\,
\end{equation}
where $F\big(v,r)=0$. There is also an event horizon \emph{H} at
\begin{equation}
w_{H}\equiv v_s-4M\left\{1+W\left(-\exp\Big[-\frac{\Delta v}{4M}-1\Big]\right)\right\},
\end{equation}
inside the trapping horizon. 

\begin{figure*}
\centering
\includegraphics[width=0.35\textwidth]{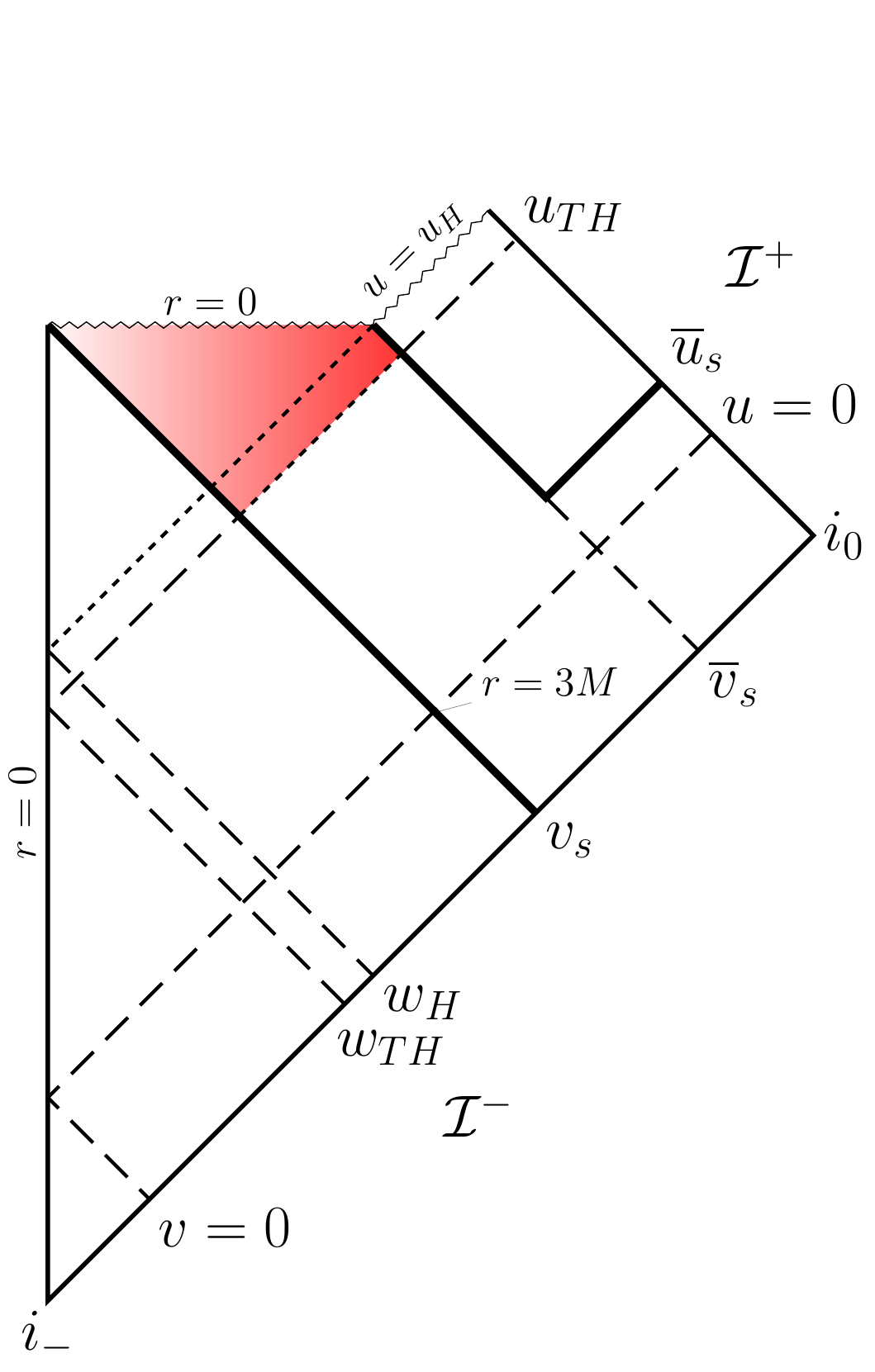} \hfill
\includegraphics[width=0.55\textwidth]{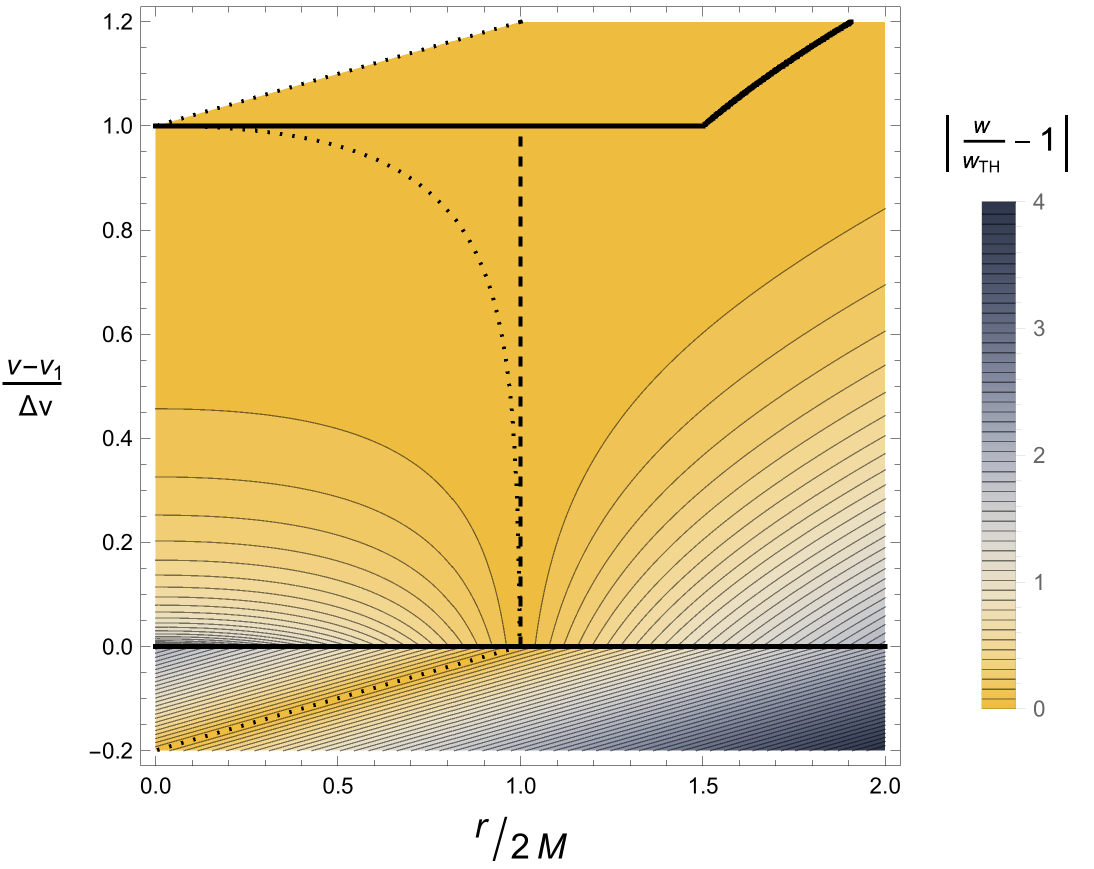}
\caption{Geometry of a singular black-hole evaporation model. Left: Carter-Penrose diagram, with the trapped region shaded. Right: outgoing null geodesics in Eddington-Finkelstein coordinates. Both diagrams show: the shells (thick line), the event horizon (thin dotted line) and the trapping horizon (thick dotted line). Note the difference between the event and trapping horizons, and the existence of (quantum) null singularity along $u=u_H$.}
\label{fig:geo_hawking}
        \end{figure*}   

The canonical map $w=w(u)$ from $\mathcal{I}^+$ to $\mathcal{I}^-$ can be easily computed and is given by
\begin{numcases}{w(u)=}
\label{eq:singular-w}
v_s-4M\left\{1+W\left(\frac{v_{s}-4M}{4M}\exp\left[-\frac{u-v_{s}+4M}{4M}\right]\right)\right\} & if $u\leq \overline{u}_s$, \\[.5em]
v_s-4M\left\{1+W\left(\frac{u_H-u-4M}{4M}\exp\left[-\frac{u-u_H+4M+\Delta v}{4M}\right]\right)\right\} \!\!\!\!\!  & if $\overline{u}_s< u\leq u_{H}$,\nonumber
\end{numcases}
where
\begin{equation}
  \overline{u}_s=\overline{v}_s-2\overline{R}_s+4M\log\left(\frac{v_s-4M}{2\overline{R}_s-4M}\right)
\end{equation}
is the retarded time of the shell-pair production event. The event horizon $H$ has retarded time
\begin{equation}
  u_{H}=\overline{u}_s+2\overline{R}_s=\overline{v}_s+4M\log\left(\frac{v_s-4M}{2\overline{R}_s-4M}\right).
  \label{uSingular}
\end{equation}

\begin{figure}
\begin{center}
\includegraphics[scale=.78]{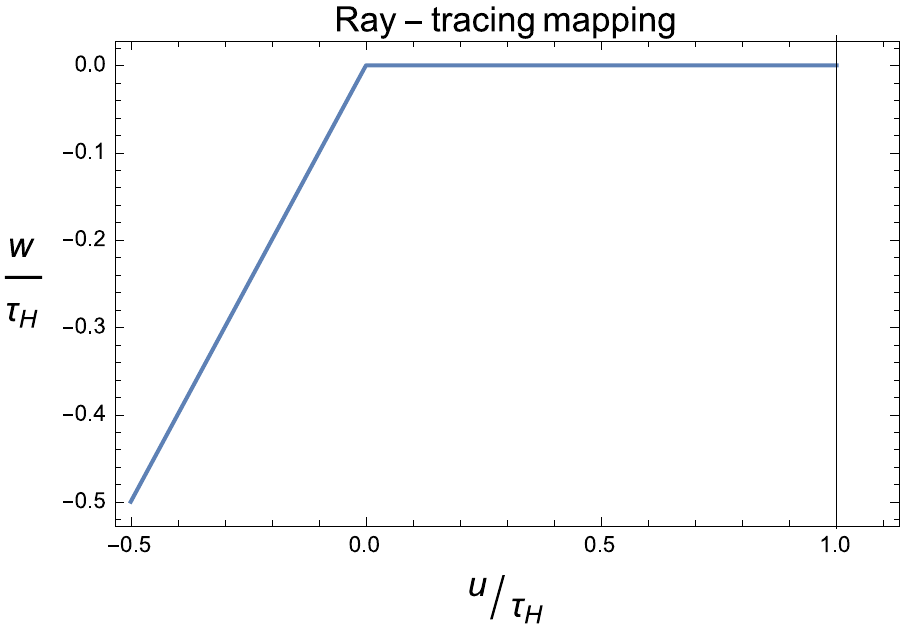}\hfill
\includegraphics[scale=.85]{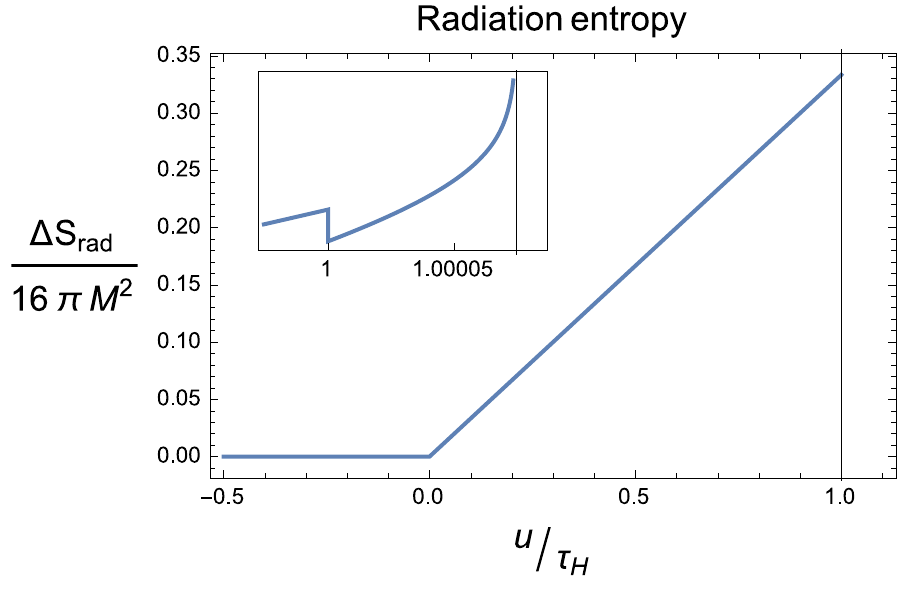}
\end{center}
\caption{Ray-tracing mapping $w=w(u)$ (left) and radiation entropy $\Delta S_{\text{rad}}(u)$ (right) in the singular evaporation model with $M=10m_P$ and $\overline{R}_s=3M$. Inset: Late-time behavior of the radiation entropy, showing the $\mathcal{O}(1)$ discontinuity and the divergence at $u=u_H$.}
\label{fig:entropy_sing}
\end{figure}

Plugging \eqref{uSingular} into the expression \eqref{Sradraytracingmapping} for the radiation entropy we find the ``Page curve'' shown in Fig.~\ref{fig:entropy_sing}. Up to the time $\overline{u}_s$, the radiation entropy grows exactly as in the Vaidya case, Eq.~(\ref{S-linear}). In particular if this evaporation phase is assumed to last a long time $\bar{u}_s=\tau_H\equiv \alpha\, M^3/m_P^2$ as in the standard Hawking evaporation scenario \cite{Hawking:1974sw}, the radiation entropy reaches the value 
\begin{equation}
\Delta S_{\text{rad}}(\overline{u}_s)_-\approx \frac{\alpha}{48}\frac{M^2}{m_P^2}\,.
\end{equation}
The entropy of the radiation up to this time matches the scaling of the Bekenstein-Hawking entropy of the black hole, i.e. $S_{\text{BH}}=A/4 \ell_P^2\sim M^2/m_P^2$. 

Then at the time $\overline{u}_s$, the entropy has a discontinuity
\begin{equation}
\left[\Delta S_{\text{rad}}(\overline{u}_s)\right]\equiv\Delta S_{\text{rad}}(\overline{u}_s)_+-\Delta S_{\text{rad}}(\overline{u}_s)_-=-\frac{1}{12}\log(\overline{R}_s/2M-1).
\end{equation}
This discontinuity is an artifact of the background metric chosen to model the evaporation process (outgoing null rays with $u<\overline{u}_s$ are more redshifted than outgoing null rays with $u>\overline{u}_s$), without any practical significance for macroscopic holes: $\left[\Delta S_{\text{rad}}(\overline{u}_s)\right]$ is of order $1$, while the entropy itself is of order $(M/m_P)^2$.

Finally at later times, when $u\to u_{H}$, the entanglement entropy of the radiation presents a divergent behavior. In particular, for $u\to u_{H}$, we find
\begin{equation}
\Delta S_{\text{rad}}(u)\sim\frac{\alpha}{48}\frac{M^2}{m_P^2}+\frac{1}{12}\log \left(\frac{4M}{u_H-u}\right).
\label{eq:Sdivergent}
\end{equation}
This naked ``entanglement singularity'' along the null ray $u=u_{H}$ is a consequence of the form of the background metric and is accompanied by a quadratic divergence in the energy flux of the radiation emitted at the last ray \cite{Hiscock:1980ze,Hiscock:1981xb}, 
\begin{equation}
F(u)\sim \frac{1}{16\pi(u_{H}-u)^2}\,.
\end{equation}
This flux has an infinite total energy. A similar lightlike singularity was dubbed a ``thunderbolt''  in \cite{Hawking:1992ti}.  As noticed in \cite{Hiscock:1980ze,Hiscock:1981xb} the pathological behavior of the flux $F(u)$ is a consequence of the assumed form of the mass function. Indeed a necessary condition for a finite total energy emitted is that the mass function goes to zero smoothly. Enforcing the latter, however, does not garantee that $\Delta S_{\text{rad}}(u)$ remains finite as $u\to u_H$; indeed we expect that $\Delta S_{\text{rad}}(u)$ will diverge in finite time in any model where a singularity meets an event horizon.\footnote{Let us emphasize, however, that truncating the quantum field at some high energy scale, e.g. the Planck scale, would regulate the divergence of both $F(u)$ and $\Delta S_{\text{rad}}(u)$. Moreover regulating the logarithmic divergence in Eq.~\eqref{eq:Sdivergent} with a Planck scale cut-off would lead to a maximum of the entropy $\Delta S_{\text{rad}}\sim\frac{\alpha}{48}\frac{M^2}{m_P^2}+\frac{1}{12}\log \big(\frac{4M}{m_P}\big)$ that still scales quadratically with the mass of the black hole.} Such null entanglement singularities would then be the manifestation of ``information loss" in singular black hole evaporation.

\subsection{Nonsingular black hole evaporation}

Next, we consider a model of \emph{nonsingular} black hole formation and evaporation with one closed trapped region. Such spacetimes, which have the same causal structure as Minkowski space (in particular, there is no event horizon) were introduced in \cite{Bardeen:1968,Frolov:1979tu,Frolov:1981mz,Stephens:1993an,Ashtekar:2005cj,Ashtekar:2008jd,Bonanno:2006eu,Hossenfelder:2009fc,Hayward:2005gi,Rovelli:2014cta,Frolov:2014jva,Bardeen:2014uaa} as a toy models for \emph{unitary black hole evaporation}.

The metric has the form (\ref{eq:F(v,r)}) with the function $F(v,r)$ given by 
\begin{numcases}{F(v,r)=}
\label{nonsingmetric}
1 & if $v < v_s$, \\[.5em]
1-\frac{2M}{r}& if $\{v_s\leq v < \overline{v}_s,\;r>\overline{R}_s\}$, \nonumber\\[.5em]
1-\frac{r^2}{\ell^2}& if $\{v_s\leq v < \overline{v}_s,\;r\leq \overline{R}_s\}$, \nonumber\\[.5em]
1-\frac{2M}{r}& if $\{v\geq \overline{v}_s ,\;r>r_0(v,\overline{w}_s)\}$, \nonumber\\[.5em]
1& if $\{v\geq \overline{v}_s ,\;0\leq r\leq r_0(v,\overline{w}_s)\}$. \nonumber
\end{numcases}
This spacetime is flat for $v<v_s$; in the slab $v_s\leq v\leq \overline{v}_s$, it consists of a Schwarzschild patch with mass $M$ for $r> R_m$ and a de Sitter patch with cosmological constant $\Lambda\equiv 1/\ell^{2}$ for $r\leq R_m$. There is no black-hole singularity in this model, only a core of Planckian curvature. The (spacelike) matching surface $R_m$ is assumed to lie within the trapped region, enclosed between the \emph{inner trapping horizon} $r=\ell$ and the \emph{outer trapping horizon} $r=2M$.  We take $R_m\equiv(2M\ell^{2}\,)^{1/3}$. For $v\geq \overline{v}_s$, the structure is the same as in (\ref{eq:hiscock-evap}): a positive and a negative mass shell originating at $r=\overline{R}_s>2M$ model the evaporation process \cite{Hiscock:1980ze,Hiscock:1981xb}.

Let us define the relevant advanced times for this metric. As before we call $w_1\equiv\overline{w}_s$ the time when the two shells modeling the evaporation process are produced, i.e. $r_0(\overline{v}_s,\overline{w}_s)=\overline{R}_s$. Moreover we define $w_2$ as the advanced time when the negative-energy shell reaches the matching surface, i.e. $r_0(\overline{v}_s,w_2)=R_m$, $w_3$ and $w_4$ as the times when the positive-energy shell reaches respectively the matching surface, $r_0(v_s,w_3)=R_m$, and the inner horizon, $r_0(v_s,w_4)=\ell$, $w_5$ as the time when $w_5=v_s$ and $w_6=\overline{v}_s$. As before we denote 
\begin{equation}
 w_{TH}\equiv v_s-4M
\end{equation}
the position of the outer trapping horizon and $\Delta v\equiv \overline{v}_s-v_s$.

In this toy model the canonical map $u\mapsto w(u)$ from $\mathcal{I}^+$ to $\mathcal{I}^-$ can be computed analytically. It shares the main features of the metrics considered in \cite{Frolov:1981mz,Hayward:2005gi,Rovelli:2014cta,Frolov:2014jva}, for which a numerical integration is needed. We find
\begin{numcases}{w(u)=}
\label{eq:nonsingular-w}
\textstyle w_{TH}-4M\,W\Big(\frac{w_{TH}}{4M}\exp[-\frac{u-w_{TH}}{4M}]\Big) & if $u<u_1$, \\[.5em]
\textstyle w_{TH}-4M\,W\Big(\frac{\overline{u}_s-u+2\overline{R}_s-4M}{4M}\exp\left[-\frac{u-\overline{u}_s-2\overline{R}_s+4M+\Delta v}{4M}\right]\Big) & if $u_1\leq u< u_2$,\nonumber\\[.5em]
\textstyle w_{TH}-4M W\Big(\frac{R_m-2M}{2M}\,\exp[{-\frac{J(u-u_6)-2R_m+4M+\Delta v}{4M}}] \Big) & if $u_2\leq u< u_3$,\nonumber\\[.5em]
\textstyle v_s+ 2\ell\coth\Big[\f{\Delta v}{2\ell} 
+\arccoth\big[\f{u-u_6}{2\ell}\big]\Big]& if $u_3\leq u< u_4$,\nonumber\\[.5em]
\textstyle v_s+ 2\ell\tanh\Big[\f{\Delta v}{2\ell} 
+\arctanh\big[\f{u-u_6}{2\ell}\big]\Big]& if $u_4\leq u< u_5$,\nonumber\\[.5em]
\textstyle \overline{v}_s+2\ell\arctanh\big[\f{u-u_6}{2\ell}\big] & if $u_5\leq u< u_6$.\nonumber\\[.5em]
\textstyle u+4M \log\Big(\frac{u-u_6+2\overline{R}_s-4M}{v_s-4M}\Big) & if $u\geq u_6$,\nonumber
\end{numcases}
where the function $J(u)$ is 
\be
J(u)\equiv 2\ell\,\big(\arccoth\big[u/2\ell\big]+\arccoth[R_m/\ell] \big)
\ee
and we defined the retarded times corresponding to $w_1,\dots,w_5$:
\begin{align}
u_1&=\overline{u}_s\,,\\
u_2&=\overline{u}_s+2(\overline{R}_s-R_m)\,,\nonumber\\
u_3&=\overline{u}_s+2\overline{R}_s-2\ell\coth\left[\f{\Delta v}{2\ell}+\arccoth\left[\frac{R_m}{\ell}\right]\right]\,,\nonumber\\
u_4&=\overline{u}_s + 2\overline{R}_s -2\ell\,,\nonumber\\
u_5&=\overline{u}_s+2\overline{R}_s-2\ell\tanh\left[\f{\Delta v}{2\ell}\right]\,,\nonumber\\
u_6&=\overline{u}_s+2\overline{R}_s.\nonumber
\end{align}
The first two lines in (\ref{eq:nonsingular-w}) coincide with (\ref{eq:singular-w}) and describe the propagation of light rays that never enter the core of the non-singular black hole, while the 3rd, 4th, 5th and 6th line describe outgoing light rays that travel through the de Sitter core to finally reach future infinity. The 7th line describes light rays that propagate in Schwarzschild space and then in flat space after the disappearance of the non-singular black hole.

\begin{figure*}
\centering
\includegraphics[height=9cm]{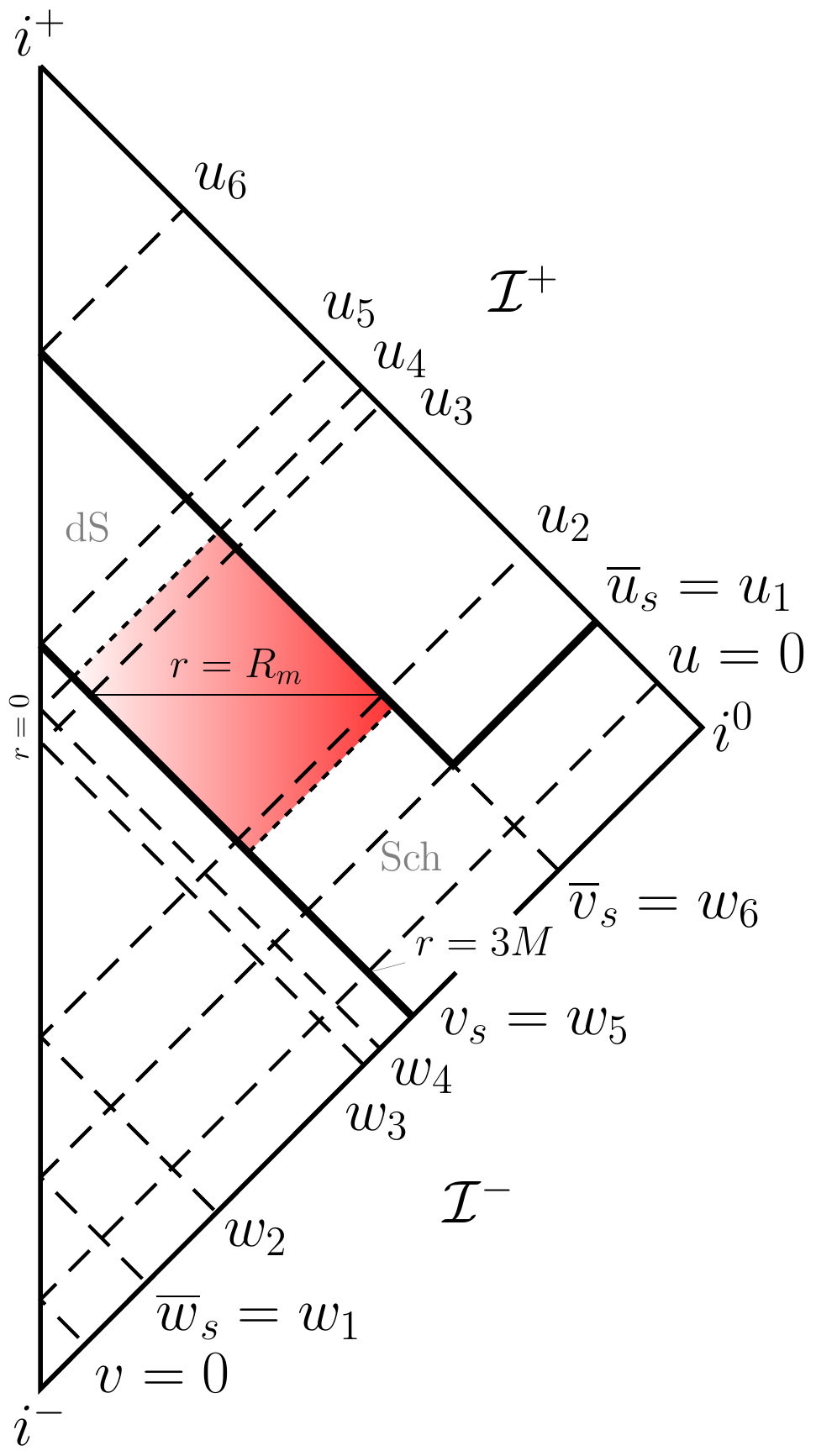}
\hfill
\includegraphics[width=0.60\textwidth]{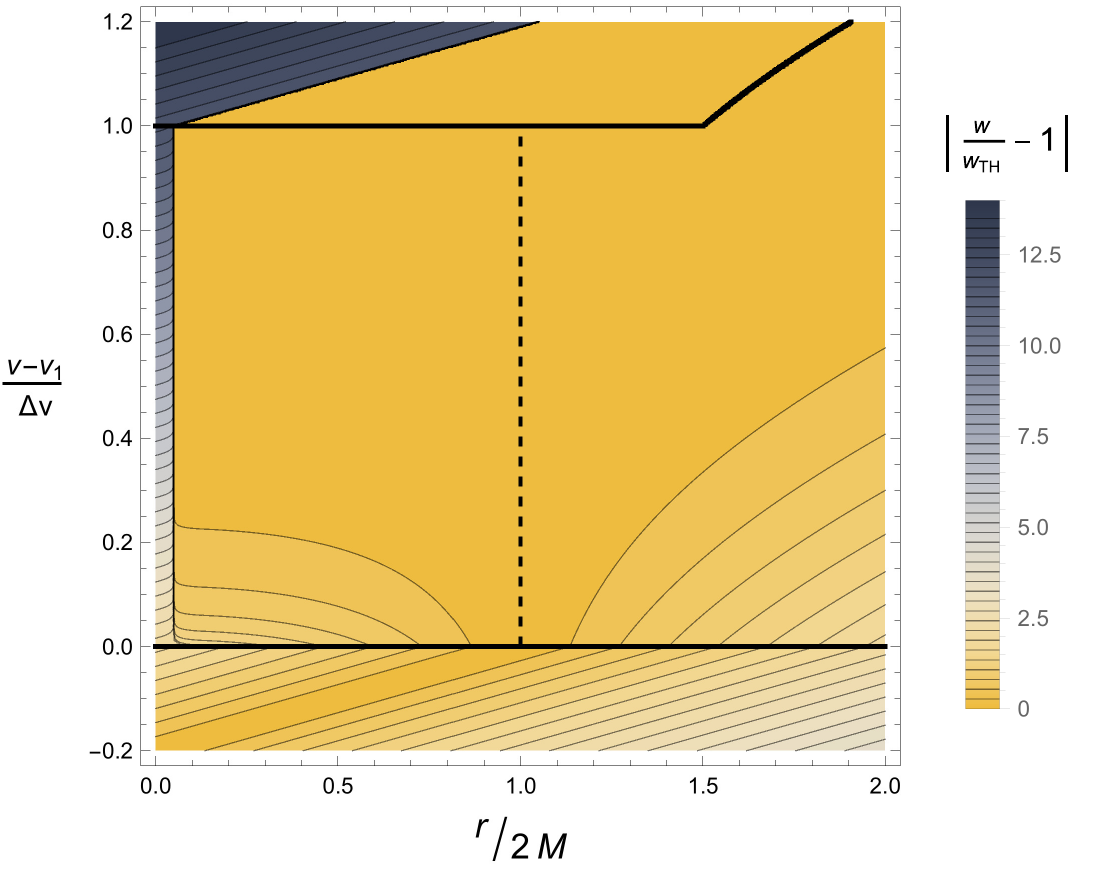}
\caption{Geometry of a non-singular black-hole evaporation model. Left: Carter-Penrose diagram, with the trapped region (shaded) and a matching surface (horizontal line) between the de Sitter core (dS) and the Schwartzschild region. Right: outgoing null geodesics in Eddington-Finkelstein coordinates, together with the shells (thick lines)  and the trapping horizon (thick dotted line).}
\label{fig:geo_reg_core}
       \end{figure*}

\begin{figure}
\begin{center}
\includegraphics[scale=.76]{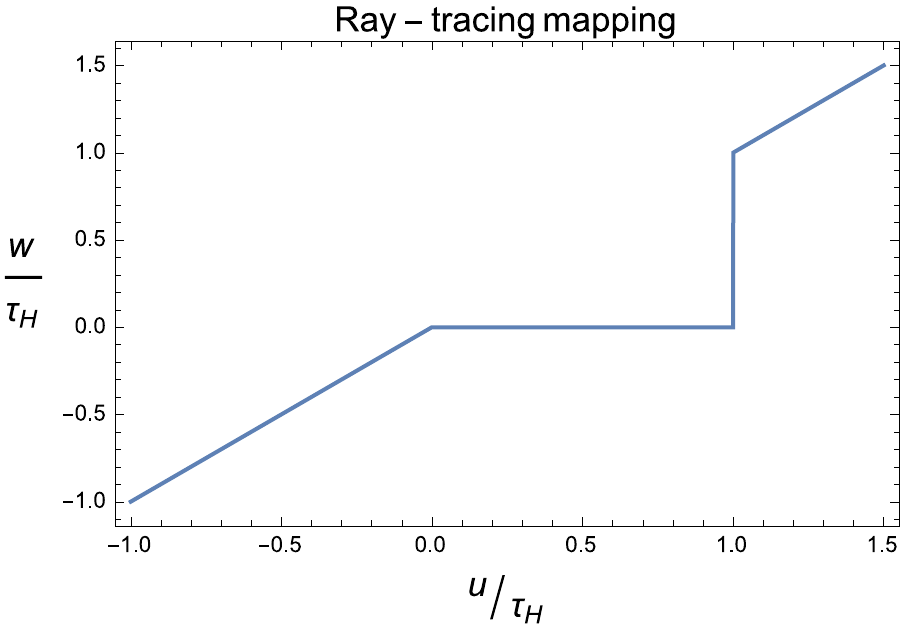}\hfill
\includegraphics[scale=.87]{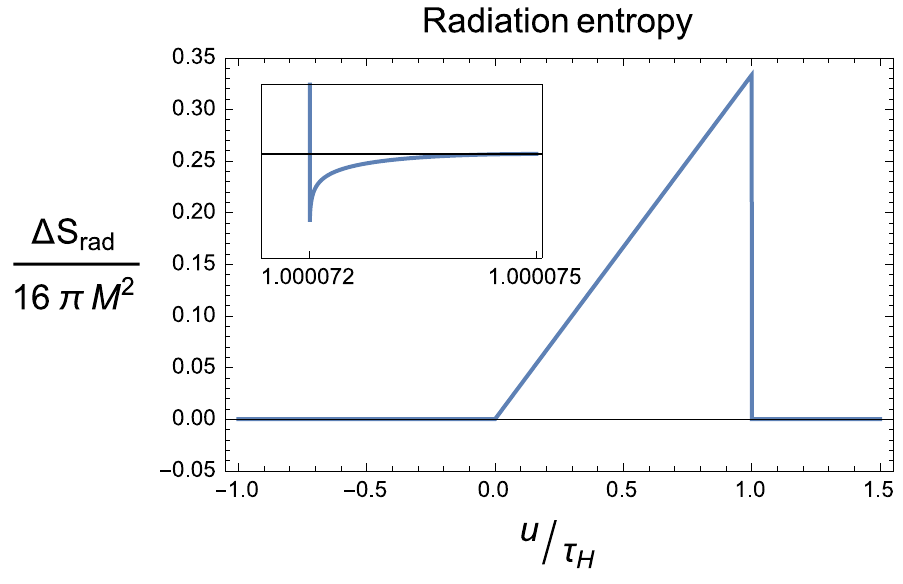}
\end{center}
\caption{Ray-tracing mapping $w=w(u)$ (left) and radiation entropy $\Delta S_{\text{rad}}(u)$ (right) in the nonsingular evaporation model, with $M=10m_P$, $\ell=M/10$ and $\overline{R}_s=3M$. Inset: Late-time behavior of the radiation entropy, showing that $\Delta S_{\text{rad}}(u)<0$ in phase C.}
\label{fig:entropy_nonsing}
\end{figure}

The entanglement entropy of the radiation emitted by the non-singular black hole can be computed using the formula derived in (\ref{Sradraytracingmapping}), see Fig.~\ref{fig:entropy_nonsing}. We distinguish four phases of the evolution of the non-singular black hole, phase $A$, $B$, $C$ and $D$.

Phase $A$ is indistinguishable from the standard Hawking evaporation phase in the spacetime studied in Sec.~\ref{sec:vaidya-shell-antishell}. This phase lasts for a retarded time $\Delta u_A$ approximately given by
\begin{equation}
\Delta u_A\equiv u_2\approx \alpha\, M^3/m_p^2\,.
\end{equation}
In this phase the entropy grows monotonically and reaches a maximum at
\begin{equation}
S_{\text{max}}\equiv  \Delta S_{\text{rad}}(u_2)\approx \frac{\alpha}{48}\frac{M^2}{m_P^2}+\frac{1}{12}\log\frac{2M}{R_m}\,.
\label{eq:Smax-nonsingular}
\end{equation}
The first term matches the scaling of the Bekenstein-Hawking entropy $S_{\text{BH}}=A/4 \ell_P^2\sim M^2/m_P^2$. The second diverges for $R_m\to 0$ and gives a small logarithimc correction to the Bekenstein-Hawking-like scaling for $R_m\equiv(2M\ell^{2}\,)^{1/3}$.

Phase $B$ corresponds to outgoing radiation that has been trapped by the null shell falling at $v_s$, it has entered de Sitter core of the black hole, and then been expelled when the black hole terminates its evaporation. This phase lasts a short retarded time
\begin{equation}
\Delta u_B\equiv u_5-u_2\approx 2R_m-2\ell\,.
\end{equation}
In this phase the entropy decreases monotonically and reaches its minimum at the negative value
\begin{equation}
S_{\text{min}}\equiv \Delta S_{\text{rad}}(u_5)\approx -\frac{\alpha}{12}\frac{M^3}{\ell\, m_P^2}\,.
\end{equation}
The fact that the \emph{excess} entanglement entropy of the radiation becomes negative means that the state is less correlated than the Minkowski vacuum state.

Phase $C$ corresponds to outgoing radiation that has entered the black hole after it formed at $v_s$ and before it disappeared completely at the time $\overline{v}_s$. This radiation traveled through the de Sitter core and is expelled at the end of the evaporation. This phase lasts a finite time 
\begin{equation}
\Delta u_C\equiv u_6-u_5\approx 2\ell\,.
\end{equation}
in which the entanglement entropy of the radiation grows monotonically from its minimum and approaches zero from below, quadratically for $u\to u_6$
\begin{equation}
\Delta S_{\text{rad}}(u)\sim -\frac{(u-u_6)^2}{48\, \ell^2}\,.
\end{equation}
At the end of phase C the black hole has disappeared and the entropy of radiation vanishes. Note however that this is not yet the end of the process.

Phase $D$ corresponds to late outgoing radiation that that has never entered the black hole as it  fell in after its disappearance at $\overline{v}_s$. Therefore this radiation travels through Schwarzschild space (for $u<\overline{u}_s$), then through flat space, and it reaches $\mathcal{I}^+$ at a time $u>u_6$, see Fig.~\ref{fig:entropy_nonsing}. The entanglement entropy for this phase is
\begin{equation}
\Delta S_{\text{rad}}(u)=-\frac{1}{12}\log\Big(1+\frac{4M}{u-u_6+2\overline{R}_s-4M}\Big)\,.
\end{equation}
In particular at the beginning of phase D the entanglement entropy is negative and equal to $\Delta S_{\text{rad}}(u_6)\simeq -0.1$ for $\overline{R}_s=3M$, while at late times it goes as 
\begin{equation}
\Delta S_{\text{rad}}(u)\sim -\frac{1}{12}\,\frac{4M}{u-u_6}
\end{equation}
and vanishes in the limit $u\to+\infty$, see Fig.~\ref{fig:entropy_nonsing}. This result is consistent with the unitary evolution of a quantum massless field from $\mathcal{I}^-$ to $\mathcal{I}^+$; in particular, as expected no information is lost in this model.

A remark on the generalized second law is in order. This law states that the entropy of a black hole $S_{BH}=A_H/4\ell_P^2$ plus the entropy matter outside the black hole never decreases in time \cite{Bekenstein:1973ur,Hawking:1974sw}. The existing proofs of this law rely on the existence of an event horizon and $A_H$ is understood as the area of a section of the event horizon at a given time \cite{Fiola:1994ir,Sorkin:1997ja,Wall:2009wm,Wall:2010cj,Wall:2011hj,Wall:2011kb}. In the model of nonsingular black hole evaporation considered here, no event horizon is present, only a trapped region bounded by a trapping horizon. Nevertheless in phase $A$, i.e. up to the time $u_2$, the evolution of the the quantum field and of the nonsingular black hole is undistinguishable from the one of a singular black hole with an event horizon. In this phase, lasting a time  $\Delta u_A\sim M^3/m_P^2$, we expect the generalized second law to hold. However after this phase the distinction between event horizon and trapping horizon becomes crucial: information eventually leaks out in the second case and the generalized entropy  $S_{BH}(u)+\Delta S_{\text{rad}}(u)$ decreases in time. Therefore our results on the vanishing of the entropy of radiation at late times indicate that there is no generalized second law for nonsingular black holes.

\subsection{Black hole to white hole tunnelling: ``black hole fireworks''}

As a last example we consider the model of bouncing black hole---or ``black hole fireworks"---proposed by Haggard and Rovelli in \cite{Haggard:2014ux}. In this scenario information is preserved in gravitational collapse by a quantum-gravitational tunnelling process, of which Hawking evaporation is only a higher-order ``dissipative" correction.

\begin{figure}
\begin{center}
\includegraphics[width=0.6\textwidth]{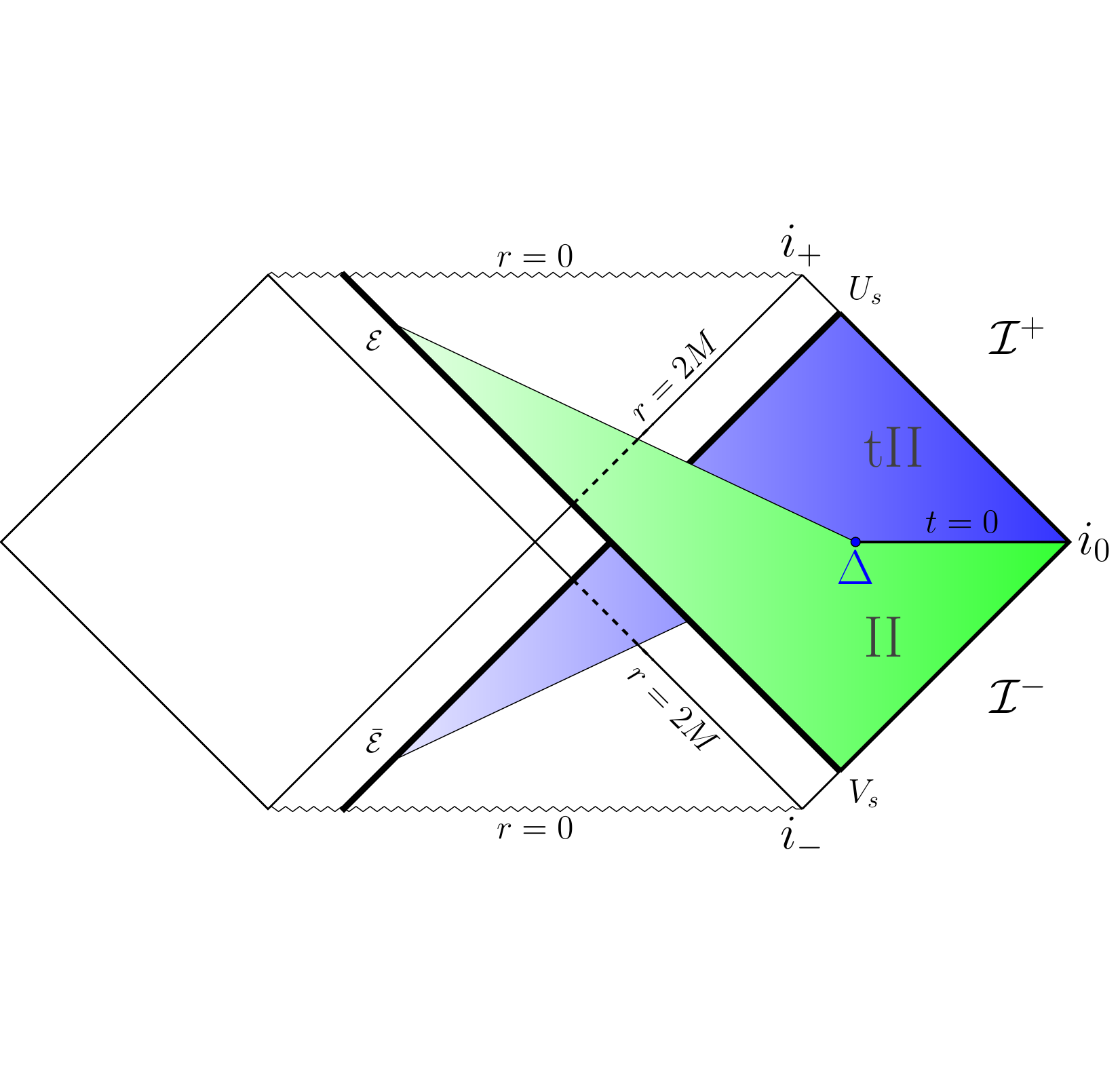} \hfill
\includegraphics[width=0.3\textwidth]{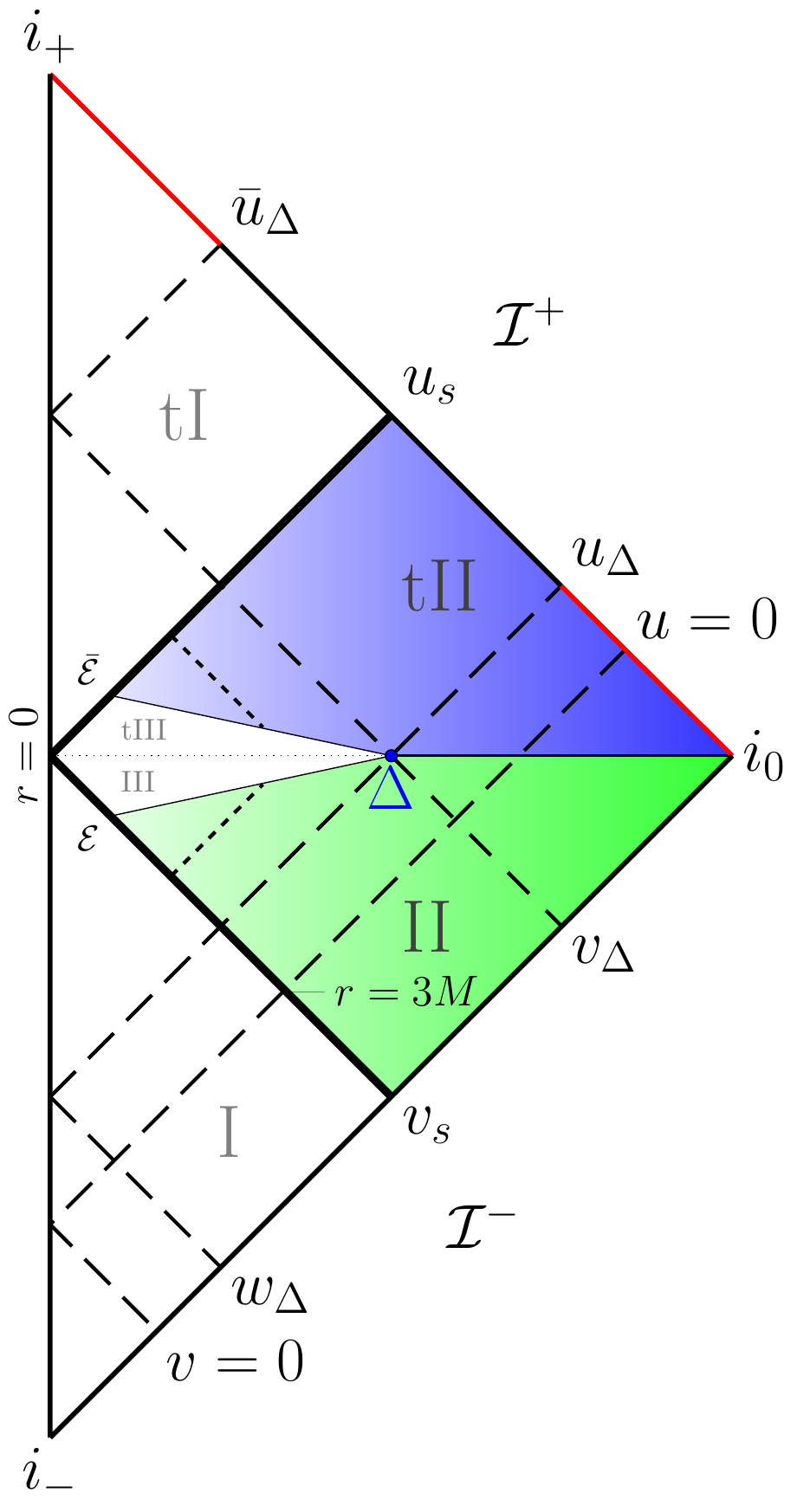}
\caption{Geometry of a time-symmetric bouncing shell. Left: Kruskal-Szekeres diagram of the extended Schwarzschild spacetime from which the model is constructed. Right: The resulting Carter-Penrose diagram of the Haggard-Rovelli fireworks spacetime.}
\label{fig:penroseBHfire}
\end{center}
\end{figure}

The corresponding spacetime is shown in Fig.~\ref{fig:penroseBHfire}. Start from the Kruskal-Szekeres diagram of an eternal black hole and pick a point $\Delta$ in the exterior region; $\Delta$ has Kruskal-Szekeres coordinates $(U_\Delta=-V_\Delta, V_\Delta)$. Choose then a null surface $V=V_s$ such that $V_\Delta > V_s$ and a point $\mathcal{E}$ on it, with coordinates $(U_\mathcal{E},V_\mathcal{E}=V_s)$. Finally pick a spacelike surface connecting $\Delta$ to $\mathcal{E}$. Call the resulting patch of Kruskal-Szekeres spacetime Region II. This region automatically determines its time-symmetric partner (Region tII) by taking the null surface $U_s = -V_s$ and the point $\overline{\mathcal{E}}$ on it, with coordinates $(V_{\overline{\mathcal{E}}}=-V_s, U_{\overline{\mathcal{E}}}=-U_\mathcal{E})$. See Fig.~\ref{fig:penroseBHfire}-Left. The Carter-Penrose diagram in Fig.~\ref{fig:penroseBHfire}-Right is then obtained by ``opening up the wings'' of Region II and tII, inserting interpolating III+tIII Regions in between, and gluing two flat Regions (I and tI) respectively along the surface $V=V_s$ and $U=U_s$.

The resulting spacetime represents the dynamics of a null infalling shell that bounces at $r=0$ and comes out as a null outgoing shell. To allow for this, geodesics must tunnel through a non-classical Region, represented by the unknown quantum Regions III and tIII. The point $\mathcal{E}$ is the point where the ingoing shell reaches Planckian density and quantum effects start to be important, while $\Delta$ is considered as the outmost boundary of the quantum regions. The spacetime is event-horizon-free, but displays a trapping and an ``anti-trapping" surface.


Here we are interested in studying the general features of the Page curve for this model. To do this, let us first observe that Region I and II can be described by the Eddington-Finkelstein coordinates $(v,r)$ and a metric of type \eqref{eq:F(v,r)} with
\begin{numcases}{F(v,r)=}
\label{eq:FrvBHfire}
1 & if $v < v_s$ \\[.5em]
1-\frac{2M}{r}& if $v \geq v_s$  and $(v,r) \in II$. \nonumber
\end{numcases} 
Here $v$ in Region II is related to the Kruskal-Szekeres $V$ by the usual relation $V\propto\exp(v/4M)$. In the same way, Region tI and tII can be described by retarded Eddington-Finkelstein coordinates $(u,r)$, where $U\propto-\exp(-u/4M)$ in Region tII, and a metric of the type
\begin{equation}
ds^2 = -F(u,r) du^2 - 2du dr.
\end{equation}
with
\begin{numcases}{F(u,r)=}
\label{eq:FrvBHfire2}
1 & if $u > u_s$, \\[.5em]
1-\frac{2M}{r}& if $u \leq u_s$ and $(u,r) \in tII$. \nonumber
\end{numcases}
As before we take the origin $u=0$ at the retarded time when the infalling shell crosses $r=3M$. The two parameters of the model are $r_\Delta>2M$ and $\Delta v \equiv v_\Delta - v_s > 0$. The retarded time $u_\Delta$ is given by
\begin{equation}
u_\Delta=v_s+\Delta v-2r_\Delta-4M\log\left(\frac{2r_\Delta-4M}{v_s-4M}\right).
\end{equation}

The canonical map $u \mapsto w(u)$ from $\mathcal{I}^+$ to $\mathcal{I}^-$ giving the entanglement entropy production can be divided in a classical phase, corresponding to the light rays that don't enter in the quantum region when traced back (red thick region on $\mathcal{I}^+$ in Fig.~\ref{fig:penroseBHfire}-Right), and the remaining quantum phase. The relevant advanced times are $u_\Delta$, that by construction gives $u_s-u_\Delta=\Delta v$, and $\overline{u}_\Delta$ defined by $w(\overline{u}_\Delta) = v_\Delta$. The two phases give us different information: the choice of the matching surface connecting $\Delta$ to $\mathcal{E}$ and of the semiclassical metric in the quantum region strongly influence the Page curve in the domain $u_{\Delta} < u < \overline{u}_{\Delta}$, while the result in the classical regime is completely insensitive to these choices and captures the general features of the model. Since the geometry of the quantum regions III and tIII remains essentially unknown, we will only compute $w(u)$ in the classical phase.  

We obtain $w(u)$ for $u \leq u_\Delta$ and $u \geq \overline{u}_\Delta$, finding
\begin{numcases}{w(u)=}
\label{eq:w(u)BHfire}
v_s-4M\left\{1+W\left[\frac{v_s-4M}{4M}\exp \left(-\frac{u-v_s+4M}{4M} \right)\right]\right\} & if $u\leq u_{\Delta}$,\\[.5em]
u  + 4M \log \frac{u - u_s - 4M}{v_s - 4M}& if $u \geq \overline{u}_{\Delta}\,$.\nonumber
\end{numcases}
where $\overline{u}_\Delta$ is given by
\begin{equation}
\overline{u}_\Delta = u_s + 4M \left\{ 1+ W\bigg[ \frac{r_\Delta-2M}{2M} \exp \left(\frac{-\Delta v+2r_\Delta-4M}{4M}\right) \bigg] \right\}\,.    
\end{equation}
The ray-tracing map at early times is the identical to the standard Vaidya case, as we expected since in this domain the path of the ray is exactly the same. At late times, on the other hand, it easy to see that $w(u)$ can be obtained from the solution at early times implementing the substitution $u-u_\Delta \leftrightarrow v_\Delta-w$ and solving for $w$.

\begin{figure}
\begin{center}
\includegraphics[scale=.75]{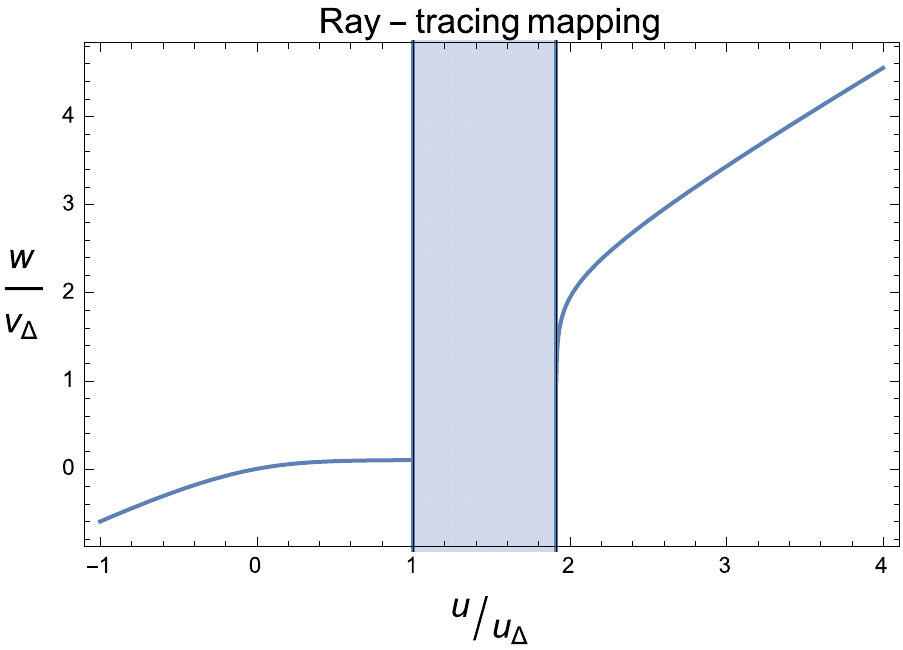}\hfill
\includegraphics[scale=.9]{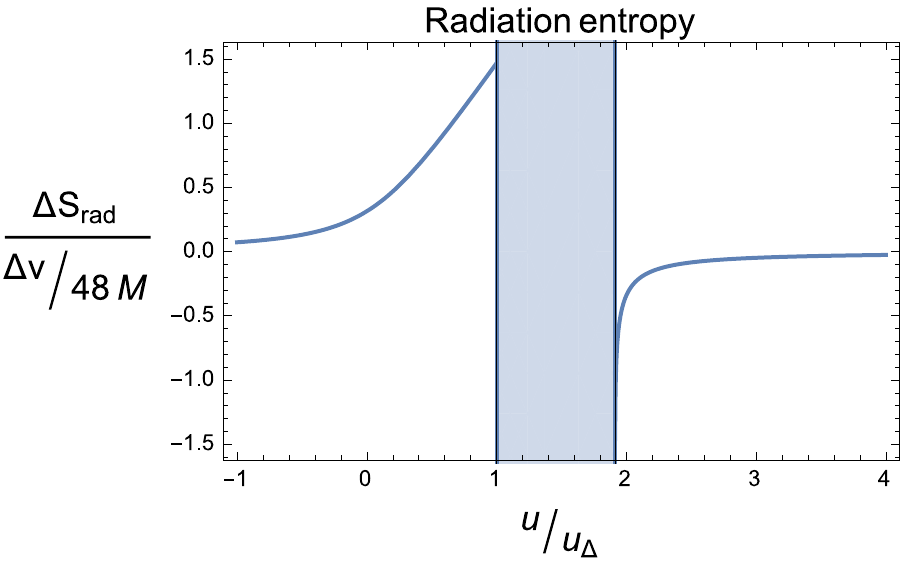}
\end{center}
\caption{Ray-tracing mapping $w=w(u)$ (left) and radiation entropy $\Delta S_{\text{rad}}(u)$ (right) in the Haggard-Rovelli `fireworks' model, with $M=10m_P$, $r_\Delta=7M/3$ and $\Delta v=1.4M$. The shaded region represents the unspecified "quantum tunnelling" phase.}
\label{fig:entropyBHfire}
\end{figure}

We can distinguish three phases in the dynamics of the radiation entropy $\Delta S_{\textrm{rad}}(u)$: phase $A$, $B$ and $C$. What we computed in equation~\eqref{eq:w(u)BHfire} is the ray-tracing map for the phases $A$ and $C$, plotted in Fig.~\ref{fig:entropyBHfire}. Exactly as before, phase $A$ is identical to the standard Hawking evaporation in a Vaidya spacetime, Eq.~\eqref{eq:entropy-vaidya}: the entropy grows monotonically and reaches a maximum at
\begin{equation}\label{Smax-fireworks}
S_{\text{max}} \equiv  \Delta S_{\text{rad}}(u_\Delta) =\frac{1}{12} \log \left( \frac{1+W \left[\frac{r_\Delta-2M}{2M} \exp \left(\frac{-\Delta v+2r_\Delta-4M}{4M}\right)\right]}{W \left[\frac{r_\Delta-2M}{2M} \exp \left(\frac{-\Delta v+2r_\Delta-4M}{4M}\right)\right]} \right)\,.
\end{equation}
In this phase, standard Hawking radiation is emitted. 
The requirement of time symmetry fixes the duration of phase $B$,
\begin{equation}
\Delta u_B\equiv\bar{u}_\Delta-u_\Delta = 
\Delta v + 4M \left\{ 1+ W\bigg[ \frac{2r_\Delta-4M}{4M} \exp \left( \frac{-\Delta v+2r_\Delta-4M}{4M}\right) \bigg] \right\}\,. 
\end{equation}
The radiation entropy in this phase depends on the geometry in the quantum region III, tIII and cannot be computed without a specific model of the effective geometry in this region.
The radiation entropy in phase $C$, i.e. for $u\geq \bar{u}_\Delta$, is given by the formula
\begin{equation}
\Delta S_{\text{rad}}(u)=-\frac{1}{12}\log\left(1+\frac{4M}{u-\bar{u}_\Delta+4M\, W \big[\frac{r_\Delta-2M}{2M} \exp \big(\frac{-\Delta v+2r_\Delta-4M}{4M}\big)\big]}\right)\,.
\end{equation}
The entropy increases monotonically from a minimum negative value to zero, see Fig.~\ref{fig:entropyBHfire}. The minimum value at the beginning of phase $C$ equals the opposite of the maximum value found in Eq. \eqref{Smax-fireworks},
\begin{equation}
S_{\text{min}}\equiv  \Delta S_{\text{rad}}(\overline{u}_\Delta)= -S_{\text{max}}.
\end{equation}
At late times the entropy approaches zero from below with the law
\begin{equation}
\Delta S_{\text{rad}}(u)\sim -\frac{1}{12}\,\frac{4M}{u-\bar{u}_\Delta}
\end{equation}
for $u\to +\infty$. As expected, the evolution of the quantum massless field from $\mathcal{I}^-$ to $\mathcal{I}^+$ is unitary and no information is lost.

We consider now two different scenarios for the scales involved in the model of bouncing black hole. The difference is in the duration of phase $A$, while we assume in both cases that the quantum region III extends outside the horizon up to a macroscopic scale\footnote{Fon instance $r_\Delta=7M/3$, as proposed in \cite{Haggard:2014ux}.} $r_\Delta\gtrsim 2M$.

In the first scenario $\Delta v=\alpha\, M^3/m_P^2$ and  phase $A$ lasts a long time $\Delta u_A\sim \tau_H \approx  \alpha\, M^3/m_P^2$ that is Hawking-like, i.e. it scales cubically with the mass of the black hole. In this case the entanglement entropy of radiation reaches a maximum $S_{\text{max}}\sim M^2/m_P^2$ at the end of phase $A$, and a minimum $S_{\text{min}}\sim-M^2/m_P^2$ and the beginning of phase $C$. Moreover, phase $B$ lasts a time $\Delta u_B\sim \alpha\, M^3/m_P^2$ and in phase $C$ the entropy reaches a value of order one, $|\Delta S_{\text{rad}}(u_f)|\sim 1$, in a time of order $\Delta u_C=u_f-\bar{u}_\Delta\sim M$. It should be noted that if phase $A$ lasts a time $\tau_H \approx  \alpha\, M^3/m_P^2$, most of the mass of the black hole is emitted in Hawking radiation and ``dissipative'' effects in the bounce cannot be neglected.

In the second scenario  $\Delta v\sim M^2/m_P$ and phase $A$ lasts a  time $\Delta u_A\sim M^2/m_P$ quadratic in the mass of the black hole. This is the scenario proposed by Haggard and Rovelli  in~\cite{Haggard:2014ux} on the basis of an estimate of cumulative quantum effects. In this case the entanglement entropy of radiation reaches a maximum $S_{\text{max}}\sim M/m_P$ at the end of phase $A$, and a minimum $S_{\text{min}}\sim -M/m_P$ and the beginning of phase $C$. Phase $B$ lasts a time $\Delta u_B \sim M^2/m_P$ and in phase $C$ the entropy becomes of order one, $|\Delta S_{\text{rad}}(u_f)|\sim 1$, in a time of order $\Delta u_C=u_f-\bar{u}_\Delta\sim M$. We emphasize that in this scenario the total energy emitted in phase $A$ in the form of Hawking radiation is small, of order $m_P$, consistently with the assumption that the process is essentially non-dissipative.\footnote{The purifying radiation emitted in phases $B$ and $C$, however, can carry away a large energy depending on the effective geometry in regions III, tIII.} The purifying phase lasts a time $\Delta u_B +\Delta u_C\sim M^2/m_P$, e.g. for a solar mass black hole a time of the order $\sim 10^{25}$ years.

\section{Conclusions}
In this paper we studied the phenomenon of entanglement entropy production during gravitational collapse and black hole evaporation. The entropy production is defined introducing a covariant regularization of the entanglement entropy, with the regulator given by the spacetime volume of the splitting region (Sec.~\ref{sec:regularization}). The main formula of the paper, Eq.~(\ref{mainformula}), is derived assuming spherical symmetry and working in the standard two-dimensional approximation: we consider only $s$-wave modes of a massless scalar field and neglect contributions from backscattering off the curved geometry. This formula allows us to give a precise, cut-off independent definition of the entanglement entropy of the exterior of a black hole \cite{Sorkin:2014kta,Bombelli:1986rw} and of the radiation that escapes from a collapsing body and reaches infinity (the Page curve) \cite{Page:1993bv}, (Sec.~\ref{sec: gravitational collapse}). We studied in detail the behavior predicted by this formula in some solvable models of gravitational collapse (Sec.~\ref{sec: Vaidya}-\ref{sec: radiation}). In particular we found that, when near-equilibrium thermodynamics and the standard description of the Hawking process of particle production apply, the entanglement entropy production matches the dynamics of the thermodynamic entropy of the radiation (Sec.~\ref{sec: Vaidya}). Remarkably the main formula holds beyond thermodynamic equilibrium and predicts interesting new features relevant for the puzzle of information loss as summarized below.

In Sec.~\ref{sec:eps-star} we studied the entropy of a quantum field on the geometry of a collapsing null shell that stops just before forming a black hole, at $r=2M+\varepsilon$. The entanglement entropy of the early radiation emitted up to a (small) finite time matches exactly the one of the radiation emitted by the incipient black hole described in Sec.~\ref{sec:vaidya-Page}. However, after a time $\sim 4M \log(2M/\varepsilon)$, the entanglement entropy drops down to zero. This phenomenon shows clearly that it is dangerous to think of the entropy as a substance: the late radiation purifies the early radiation and lowers the entropy instead of increasing it.\footnote{In the model considered the purifying radiation is emitted instantaneously at the retarded time $u_L$. The phenomenon persists when the halting of the shell is not instantaneous but takes a short finite time.} This phenomenon reproduces qualitatively the behavior described in  \cite{Page:1993bv} for the evolution of the radiation entropy in a unitary process.

In the presence of an event horizon one does not expect the entanglement entropy of the radiation to ever go back to zero: modes of the field at infinity and across the event horizon are correlated and, when the black hole evaporates completely, the information stored in this correlations is lost for observers at infinity. In Sec.~\ref{sec:vaidya-shell-antishell} we studied the entanglement entropy of the radiation emitted by an evaporating black hole with an event horizon. We considered a solvable model of evaporation consisting in the production of a single pair of shells  \cite{Hiscock:1980ze,Hiscock:1981xb}: one of positive energy radiating away all the mass of the black hole at the time $\overline{u}_s$, the other of negative energy that makes the black hole disappear. The entropy of the radiation emitted up to the time $\overline{u}_s$ matches the scaling of the Bekenstein-Hawking entropy of the black hole, $S(\overline{u}_s)\sim M^2/m_P^2$. After this time the entropy keeps increasing. We notice that this model presents a pathology right before the last ray $u_{H}$: the total energy emitted in a finite time is divergent. This thunderbolt appears together with a divergence of the entanglement entropy. While the former can be cured by a smoothly decreasing mass functions, the divergence of the entanglement entropy at the event horizon appears to be generic.

We also considered a solvable model of non-singular black hole formation and evaporation. In this model the core of the black hole consists of a de Sitter region of Planckian curvature. As a result there is no event horizon but only a closed trapped region. We showed that in this model the entanglement entropy of the radiation emitted grows monotonically up to a finite time $\overline{u}_s$ and matches exactly the curve obtained for a more standard singular black hole. This is however the first of four phases: after having reached a maximum, the entropy decreases to negative values, and then increases approaching zero from below. As expected, because of the absence of an event horizon, the entropy at late times goes back to zero. The same qualitative behavior---though with radically different time scales---was found in the Haggard-Rovelli ``black hole fireworks" model. In both cases, evolution can go through a phase where the radiation is less correlated than the vacuum and $\Delta S_{\text{rad}}(u)$ is negative. The extent to which this behavior is consistent with energy conservation, namely the requirement that the energy radiated matches the mass loss of the black hole, will be discussed elsewhere.

\section*{Acknowledgements}
We thank Abhay Ashtekar, Ted Jacobson and Carlo Rovelli for useful discussions. TDL is grateful to the Perimeter Institute for hospitality within the Summer Undergraduate Student Program. Research at the Perimeter Institute is supported in part by the Government of Canada through Industry Canada and by the Province of Ontario through the Ministry of Research and Innovation. This work was supported in part by the NSF grant PHY-1404204.


\providecommand{\href}[2]{#2}\begingroup\raggedright\endgroup

\end{document}